\documentclass[draft]{agujournal2019}
\usepackage{url} 
\usepackage{lineno}
\usepackage{graphicx}
\usepackage{mathtools,amsmath,amssymb}
\usepackage{mathrsfs}
\usepackage{amsfonts}
\usepackage{bm}
\usepackage{tabu}
\usepackage[inline]{trackchanges} 
\usepackage{CJKutf8}
\usepackage{soul}
\usepackage{lmodern} 

\newcommand{\vx}{\bm{x}}

\newcommand{\ve}{\bm{\eta}}

\newcommand{\mat} [1]{\pmb{\mathsf{#1}}}
\journalname{JGR: Oceans}

\begin{document}

%
%


\title{Dynamic-Mode Decomposition of Geostrophically Balanced Motions from SWOT Altimetry}

%
%




\authors{Takaya Uchida\,(\begin{CJK}{UTF8}{min}内田貴也\end{CJK})\affil{1,2}\thanks{Will be moving to the \textit{Climate Dynamics Laboratory, Moscow Institute of Physics and Technology, Russia}.},
Badarvada Yadidya\affil{3},
Karl E.~Lapo\affil{4},
Xiaobiao Xu\affil{1},
Jeffrey J.~Early\affil{5},
Brian K.~Arbic\affil{3},
Dimitris Menemenlis\affil{6},
Luna Hiron\affil{1},
Eric P.~Chassignet\affil{1,7},
Jay F.~Shriver\affil{8},
and Maarten C.~Buijsman\affil{9}
}


\affiliation{1}{Center for Ocean-Atmospheric Prediction Studies, Florida State University, Florida, USA}
\affiliation{2}{Universit\'e Grenoble Alpes, CNRS, INRAE, IRD, Grenoble INP, Institut des G\'eosciences de l’Environnement, Grenoble, France}
\affiliation{3}{Department of Earth and Environmental Sciences, University of Michigan, Michigan, USA}
\affiliation{4}{Department of Atmospheric and Cryospheric Sciences (ACINN), Universit\"at Innsbruck, Innsbruck, Austria}
\affiliation{5}{NorthWest Research Associates, Washington, USA}
\affiliation{6}{Jet Propulsion Laboratory, National Aeronautics and Space Administration, California, USA}
\affiliation{7}{Department of Earth,\,Ocean and Atmospheric Science, Florida State University, Florida, USA}
\affiliation{8}{Naval Research Laboratory, Stennis Space Center, Mississippi, USA}
\affiliation{9}{School of Ocean Science and Engineering, University of Southern Mississippi, Mississippi, USA}




\correspondingauthor{T. Uchida}{takachanbo@gmail.com}




\begin{keypoints}
\item Dynamic-mode decomposition (DMD) is applied for the first time to sea-surface height (SSH) fields.
\item DMD extracts the sub-inertial signals from SSH that has an imprint of internal waves.
\item Slowly varying DMD spatial modes can be used to isolate geostrophically balanced motions.
\end{keypoints}

%
%

%
%


\begin{abstract}
The decomposition of oceanic flow into its balanced and unbalanced motions carries theoretical and practical significance for the oceanographic community. These two motions have distinct dynamical characteristics and affect the transport of tracers differently from one another. The launch of the Surface Water and Ocean Topography (SWOT) satellite provides a prime opportunity to diagnose the surface balanced and unbalanced motions on a global scale at an unprecedented spatial resolution. Here, we apply dynamic-mode decomposition (DMD), a linear-algebraic data-driven method, to tidally-forced idealized and realistic numerical simulations and one-day-repeat SWOT observations of sea-surface height (SSH) in the separated Gulf Stream. DMD is able to separate out the spatial modes associated with sub-inertial periods from super-inertial periods. The sub-inertial modes of DMD can be used to extract geostrophically balanced motions from SSH fields, which have an imprint of internal tides and gravity waves. 
We utilize the statistical relation between relative vorticity and strain rate as the metric to gauge the extraction of geostrophy.
\end{abstract}

\section*{Plain Language Summary}
Observations of the global ocean surface are now done routinely by satellites. One of the key variables in describing the oceanic state is sea-surface height (SSH), i.e.,~elevations of the sea surface. For those who enjoy marine sports, it is well appreciated that the ocean surface is teeming with waves and currents. Similar to the density interface between the ocean and atmosphere, there are waves beneath the surface at density interfaces within the ocean. Both surface- and internal-wave signals imprint onto SSH. In order to extract information on oceanic currents from SSH (e.g.,~flow direction and speed), it is necessary to remove the signal of surface and internal waves since waves and currents are generally not physically related to each other on the same scales in space and time. Namely, waves tend to propagate much faster and have smaller spatial scales than the currents. Here, we implement a method based on linear algebra, which is able to capture the slowly varying residual signals from the waves.

%
%

%


%
%
%
%

\section{Introduction}
With the launch of the Surface Water and Ocean Topography (SWOT) satellite, there is great interest within the oceanographic community to extract surface velocity information from the new altimetry observations with $\mathcal{O}(1\,\text{km})$ spatial resolution \cite{dibarboure2024blend}. The fact, however, that the observed altimetry is a superposed signal of geostrophic turbulence and waves complicates the problem \cite<e.g.>[]{richman2012inferring,savage2017frequency,le2023regional,xiao2023reconstruction,Maingonnat_2024}. While geostrophy is one of the most simple and practical balances that relates sea-surface height (SSH) gradients to velocity, horizontal gradients of unfiltered SSH observations are contaminated by high-frequency balanced motions and unbalanced motions such as flows with Rossby numbers on the order of unity and larger and internal wave signals \cite{Torres2018,mcwilliams2019survey,mcwilliams2021oceanic}.

One work around has been to exploit the fact that submesoscale dynamics and unbalanced motions are associated with smaller spatial scales and shorter time scales than geostrophic eddies. Namely, filtering the SSH and/or momentum fields by band-pass filters in the wavenumber and frequency domain \cite{Wang_2023,jones2023using,bakhoday2024impact}.
A limitation of this approach is that Fourier transforms require the data to be periodic and have relatively high resolution in space-time in order to avoid aliasing. Another popular method for modal decomposition, empirical orthogonal function (EOF), is excellent at extracting spatial modes of the data but decouples the space-time information \cite{uchida2021ensemble}; the EOF spatial modes are unaware of the temporal phase information. Additionally, decomposition methods based only on spatial information do not remove internal waves that have wavelengths comparable to the local Rossby radii of deformation \cite<>[]{cao2021submesoscale}. 
Lagrangian filtering, on the other hand, requires direct knowledge of the momentum fields themselves \cite{shakespeare2021new,wang2023simple,jones2023using,baker2024lagrangian,minz2024exponential}, which SSH observations do not directly provide.

Here, we implement a relatively novel data-driven method coined as dynamic-mode decomposition (DMD) that decomposes the data into spatial modes while retaining the phase (i.e.,~growing, decaying and/or oscillating in time) information associated with each mode \cite{kutz2016dynamic}. Conceptually, it can be thought of as applying the band-pass filter in the real space-time domain (instead of the wavenumber-frequency domain). 
Or, it can be thought of as EOF spatial modes associated with temporal phase information.
DMDs have been widely adopted in the broader field of fluid mechanics \cite{schmid2022dynamic,baddoo2023physics,brunton2024promising}, plasma physics and geomagnetics \cite{chi2023extracting,kutz2024shallow}, neuroscience \cite{brunton2016extracting}, and epidemiology \cite{proctor2015discovering}.
In the context of SWOT, our quest is to decompose the slowly varying spatial modes in first-order balance with Earth's rotation and vertical stratification from the fast unbalanced modes associated with internal gravity waves (IGWs) given the observed SSH fields.
As we shall see, DMD is capable of separating out the slow (sub-inertial) component in SSH without the requirement of periodicity and will allow us to diagnose geostrophy from it. 
We will demonstrate that this method is an effective approach to isolate geostrophic motions in idealized and realistic high-resolution ocean simulations with IGWs, and one-day-repeat SWOT tracks.

The paper is organized in a way to apply multi-resolution coherent spatiotemporal scale separation (mrCOSTS), a variant of DMD, to flows from idealized configurations to increasing levels of complexity and realism.
We briefly introduce the math behind mrCOSTS and the SSH dataset from idealized and realistic tidally-forced submesoscale-permitting simulations in the section below. We present our results in Section~\ref{sec:res}. 
Section~\ref{sec:conc} ends with a Discussion on SWOT.

\section{Method and Data}
\label{sec:method}
\subsection{Multi-Resolution Coherent Spatiotemporal Scale Separation}
At the basic level, dynamic-mode decomposition (DMD) is a method that seeks a locally linear dynamical system \cite{kutz2016dynamic}
\begin{equation}
    \label{eq:cont}
    \frac{d}{dt}\ve = \mathcal{A}\ve \,,
\end{equation}
where $\mathcal{A}$ is a linear operator and approximately encapsulates all physical processes responsible for the system to step forward in time.
In discrete form, this can be recasted as 
\begin{equation}
    \label{eq:1D}
    \ve_{n} = \mat{A}\ve_{n-1} = \mat{A}^n\ve_0\,,
\end{equation}
where $\mat{A} = \exp{(\mathcal{A}\Delta t)}$ and $n=1,2,...\,$ is the time step.
$\Delta t$ is the time between the time steps when discretizing \eqref{eq:cont}.
The solution to this can be expressed by the eigenvalues $\lambda_n$ and eigenvectors $\boldsymbol{\psi}$ of the discrete-time map $\mat{A}$
\begin{equation}
    \label{eq:base}
    \ve_{n} = \sum_{j=1}^r\boldsymbol{\psi}_j \lambda_{n,j} \boldsymbol{b}_j\,,
\end{equation}
where $\boldsymbol{b}_j$ are the coordinates of the initial state $\ve_0$ in the eigenvector basis, and $r$ is the rank of singular-value decomposition (SVD) of $\mat{A}$.
Equation~\eqref{eq:1D} can be expanded without a loss of generality as
\begin{subequations}
\begin{align}
    \mat{H} &= 
    \begin{bmatrix}
        |     & |     &        & |\\
        \ve_0 & \ve_1 & \cdots & \ve_{n-1}\\
        |     & |     &        & |
    \end{bmatrix}\,, \\
    \mat{H}' &= 
    \begin{bmatrix}
        |     & |     &        & |\\
        \ve_1 & \ve_2 & \cdots & \ve_{n}\\
        |     & |     &        & |
    \end{bmatrix}\,,
\end{align}
\end{subequations}
where $\mat{H}$ and $\mat{H}'$ are shifted by one time step.
The DMD algorithm produces a low-rank eigen decomposition \eqref{eq:base} of matrix $\mat{A}$ that optimally minimizes the Frobenius norm \cite{askham2018variable}
\begin{equation}
    \label{eq:A}
    ||\mat{H}' - \mat{A}\mat{H}||_F\,.
\end{equation}
By rewriting $\omega_j = \ln{(\lambda_j)}/\Delta t$, the approximate solution for all future times can be predicted as
\begin{equation}
    \label{eq:basicdmd}
    \ve(t,\vx) \approx \sum_{j=1}^r \boldsymbol{\psi}_j(\vx) \exp{(\omega_j t)} \boldsymbol{b}_j = \mat{\Psi}\exp{(\boldsymbol{\Omega} t)}\mat{b}\,.
\end{equation}
The real part of $\omega_j$, $\text{Re}[\omega_j]$ gives growing or decaying modes in time while the imaginary part $\text{Im}[\omega_j]$ corresponds to oscillating modes.
Equation~\eqref{eq:basicdmd} may look similar to EOFs, viz.
\begin{equation}
    \label{eq:eof}
    \ve(t,\vx) \approx \sum_{j=1}^M \boldsymbol{a}_j(t)\boldsymbol{\phi}_j(\vx)\,,
\end{equation}
where $\boldsymbol{a}$ is the principle components and $\boldsymbol{\phi}$ is the EOF spatial modes \cite<cf.>[]{uchida2021ensemble}. The difference is that the spatial modes are decoupled from the temporal phase information in EOFs while the two are interlinked in DMDs.

In practice, we employ a variant of DMD, viz.~multi-resolution coherent spatiotemporal scale separation (mrCOSTS), 
a method which was originally proposed by \citeA{Dylewsky_2019} and advanced by \citeA{lapo2024multi}, to deal with datasets comprising of multi-scale non-linear dynamics by iteratively applying DMD over the entire dataset.
For each decomposition level a sliding window of fixed length in time is applied to the data and a DMD model is fit, resulting in a collection of DMD models for each window. These DMD models are categorized into a poorly resolved low-frequency component and better resolved high-frequency components, called the local-scale separation. 
The high- and low-frequencies discovered are in reference to the window length.
The low-frequency component is used as input to the next decomposition level with a larger window size.
Namely, the highest frequency components are extracted at each level.
This local-scale process is iterated over the the number of \textit{a priori} decomposition levels prescribed by the user.
Upon completion of the local scale a global-scale separation is performed, which captures leaked frequency components between decomposition levels.
The local- and global-scale separation are achieved by applying the $k$-means clustering to the temporal dynamics of the collection of DMD models \cite<see also>[their Fig.~1]{scikit-learn,lapo2024multi}.

Due to subtracting out the mean within each window, mrCOSTS is especially amenable to diagnosing fluid flows as the decomposition approximates the high-frequency fluctuations of a Reynold's averaged flow at each decomposition level. 
The resulting decomposition identifies discrete bands of coherent spatiotemporal modes. The scale-separated bands are denoted $\mathcal{G}_{1}, \mathcal{G}_{2}, \dots, \mathcal{G}_{p}$ where the subscript $p$ indexes the scale-separated bands. Each band can then be used to reconstruct the contribution of $\mathcal{G}_{p}$ to the original data, $\breve{\ve}_p(t)$,
\begin{equation}
    \breve{\ve}_p(t,\vx)=\sum_{k=1}^N\sum_{(j, \ell) \in \mathcal{G}_p} \boldsymbol{\psi}^k_{j,\ell}(\vx) \exp{(\omega^k_{j,\ell}t)} \boldsymbol{b}^k_{j,\ell} \label{eq:band-reconstruction}\,. 
\end{equation}
The subscript $\ell$ denotes the decomposition level and $j$ to index the DMD eigenvalue $\omega$ and eigenvector $\boldsymbol{\psi}$ pairs up to rank $r$ specific to the $\ell^\text{th}$ level.
The superscript $k$ is used to index the data windows so that snapshots belonging to the $k^\text{th}$ window are approximated by the decomposition.
We use $\breve{\ve}_p(t)$ to indicate an approximation of the original input signal per band $\ve_p(t)$. 
Summing over a subset of the bands, $p$, allows one to reconstruct a slow and fast component of the data.
The mrCOSTS reconstruction of the original data is achieved by summing up over all bands $\ve(t) \approx\sum_p \breve{\ve}_p(t)$.

MrCOSTS can provide a robust scale separation for a range of hyperparameters, often requiring little-to-no tuning. 
The most relevant hyperparameters are the length of the window used at each decomposition level, the SVD rank of the DMD fit at each level, any constraints on the eigenvalue solutions and eigenvalues themselves. 
We refer the reader to \citeA{lapo2024multi} and \citeA<>[their online tutorial \url{https://github.com/PyDMD/PyDMD/tree/master/tutorials/tutorial20}]{ichinaga2024pydmd} for further details regarding the implementation and user guide on mrCOSTS.


\subsection{Idealized Wave-Vortex Simulation}
\begin{figure}
    \centering
    \includegraphics[width=0.85\linewidth]{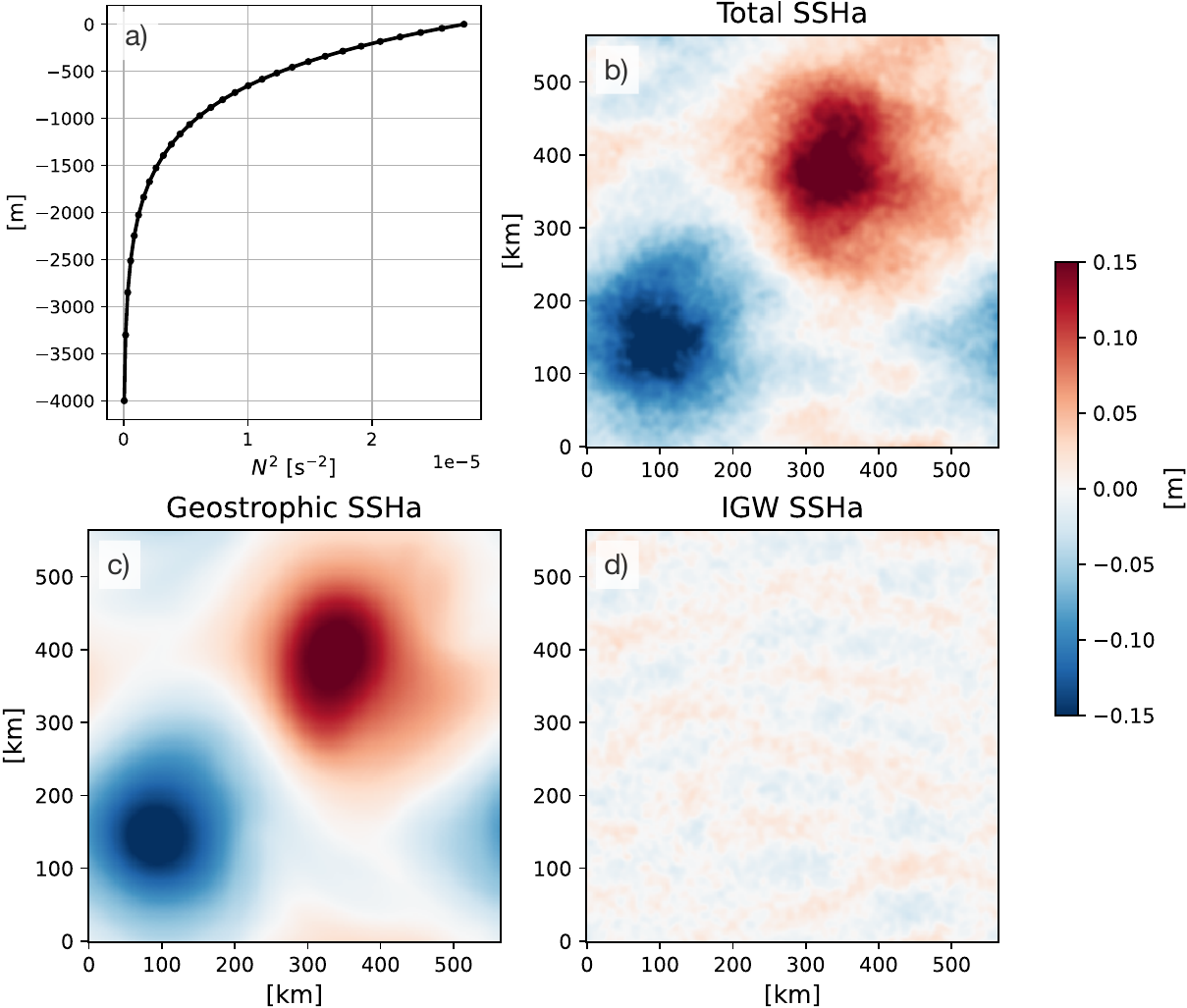}
    \caption{Buoyancy frequency ($N^2$) with an exponential vertical profile (a) and snapshots of the total SSHa and its geostrophic and IGW components at an arbitrary time step from the doubly-periodic wave-vortex simulation (b\,-\,d). 
    $N^2$ is kept stationary throughout the simulation.}
    \label{fig:w-v_SSH}
\end{figure}

As was demonstrated by \citeA{early2021generalized}, an unambiguous decomposition between linear waves and geostrophic motions can be made under flat-bottom boundary conditions. The eigenmodes from the decomposition \cite{early2024available} form a spectral basis for the wave-vortex model \cite{early2024wvm}, which then solves the equations of motion for a doubly-periodic rotating non-hydrostatic Boussinesq fluid with arbitrary stratification. At each instant in time the complete state of the fluid is decomposed into geostrophic and wave modes, while the nonlinear time steps flux energy between modes. For the simulation considered here, a single pair of dipolar geostrophic eddies were spun up from a baroclinically unstable flow, after which inertial oscillations and a narrow band of semidiurnal  internal gravity waves (IGWs) were prescribed as a forcing. The resulting steady-state simulation maintains a background IGW field following the Garrett-Munk spectrum \cite{garrett1975space} with an amplitude close to the level of SWOT instrumental noise.
The geostrophic and IGW fields are then allowed to freely evolve and interact with each other in time.
This configuration is ideal to test mrCOSTS as we will know the exact initial frequencies of the waves. 
Furthermore, because the wave-vortex model exactly isolates the geostrophic component at each time step, this will be treated as our target for mrCOSTS to extract from the total SSH anomaly (SSHa). The prescribed vertical stratification and snapshots of the spun-up SSHa of the geostrophic and IGW component at an arbitrary time step is documented in Fig.~\ref{fig:w-v_SSH}. An animation of the spun-up fields of relative vorticity normalized by the Coriolis frequency is provided in the Supporting Information (dmd-movie.mp4) where the local Rossby numbers are small ($\text{Ro} = \zeta/f \sim \mathcal{O}(0.1)$).

Given that we know \textit{a priori} that the flow consists of a single pair of geostrophic eddies and IGWs with distinct frequencies from each other, we applied mrCOSTS in two levels with the window lengths of $[1, 8]$\,days respectively; this splits the SSH anomaly (SSHa) fields into high- and low-frequency DMD modes about the window lengths during each iteration. 
In other words, the number of iterations for the local-scale separation here was prescribed as two ($N=2$ in \eqref{eq:band-reconstruction}).
The first window length was chosen to be diurnal and the second window length is the characteristic time scale of geostrophic eddies \cite<cf.>[their Fig.~3]{Torres2018}.
The model outputs were saved every 30~minutes but hourly resolution was used to construct $\mat{H}$ and $\mat{H}'$.
Conceptually, for a one-day (24-hour) and eight-day window, mrCOSTS fits 24 and 192 data points respectively in time for data with hourly resolution ($\Delta t = 1$~hours). 
Namely, the number of data points per window depends on the temporal resolution of the data used. 
The window is then slid in time to go through the entire dataset in a manner similar to how one would take the running mean. The ranks of SVD were set as $[8, 18]$, which need to be smaller than the number of data points within each window, i.e., [24, 192] respectively.
Increasing the ranks generally leads to mrCOSTS finding more modes, $\boldsymbol{\psi}^k_{j,\ell}$, but given the simplicity of the flow, we have kept it minimalistic.

Figure~\ref{fig:hist}a shows the probability density function (PDF) of frequencies associated with spatially coherent modes discovered by mrCOSTS and the frequency spectrum of SSHa over the duration of 72~days; periodograms were taken every $\sim 150$\,km and then spatially averaged to construct the spectrum. We see that the SSHa fields contain a signal of IGWs with a peak around the diurnal and semidiurnal frequencies.

\subsection{Tidally-Forced Submesoscale-Permitting North Atlantic Simulation}
We take the hourly SSHa snapshot outputs from an atmospherically and tidally forced $1/50^\circ$ North Atlantic simulation using the HYbrid Coordinate Ocean Model \cite<HYCOM50;>[]{xu2022spatial}; data are publicly available via the Open Storage Network, a cloud storage service operated by the National Science Foundation \cite<NSF;>[]{uchida2022cloud}.
HYCOM50 was spun up for 15 years from the U.S. Navy's Global Ocean Climatological (GDEM) state of rest and forced with the climatological European Centre for Medium-Range Weather Forecasts (ECMWF) Reanalysis ERA-40. Additionally, three-hourly wind anomalies from the Navy Operational Global Atmospheric Prediction System (NOGAPS) were prescribed with absolute wind stress.
Eight tidal constituents were included (K$_1$, O$_1$, P$_1$, Q$_1$, M$_2$, S$_2$, N$_2$, and K$_2$).
The three months of August\,--\,October (ASO) in year~19 is used in our analyses below.
Further details on the model configuration can be found in \citeA<>[]{xu2022spatial}.

In constructing $\mat{H}$ and $\mat{H}'$, we sub-sampled the SSHa fields every three and 12~hours in the separated Gulf Stream ($\Delta t = 3$ and $12$~hours; Fig.~\ref{fig:SSH}a), a region overlapping with a SWOT crossover (Fig.~\ref{fig:swot}). The temporal sub-sampling mimics observations where high resolution in time is not obtainable. The spatial mean was removed from each snapshot and the fields were further spatially smoothed by applying a Gaussian filter with the standard deviation of 10\,km using the {\tt gcm-filters} Python package \cite{grooms2021diffusion}; we do not expect perturbations on scales smaller than this to be in geostrophic balance \cite{pedlosky1984equations,Vallis:2006aa} and some spatial filtering is justified to compensate for the lack of temporal resolution. The latitude-longitude dimensions were flattened so as to feed mrCOSTS two-dimensional fields in space-time.

We applied mrCOSTS in six levels ($N = 6$) with each level splitting the SSH anomaly (SSHa) fields into high- and low-frequency modes about the window lengths of $[1, 2, 3, 4, 8, 30]$\,days respectively. 
The first four window lengths were chosen to be close to tidal periods and their harmonics, the fifth window length is the characteristic time scale of geostrophic eddies, and the sixth window length has a monthly time scale.
When the data were sub-sampled 12 hourly, we applied mrCOSTS in four levels ($N = 4$) with each window length corresponding to $[3, 4, 8, 30]$\,days.
The ranks of SVD were set as $[6, 6, 6, 8, 8, 10]$ in the three-hourly case for each level and $[4, 4, 6, 10]$ in the 12-hourly case. 
Note that shortening the window lengths will potentially lead to discovering bands with higher frequencies but they must be long enough to allow for the least-squares fit \eqref{eq:A} and SVD to converge. 

\begin{figure}[ht]
\centering
 \includegraphics[width=0.9\linewidth]{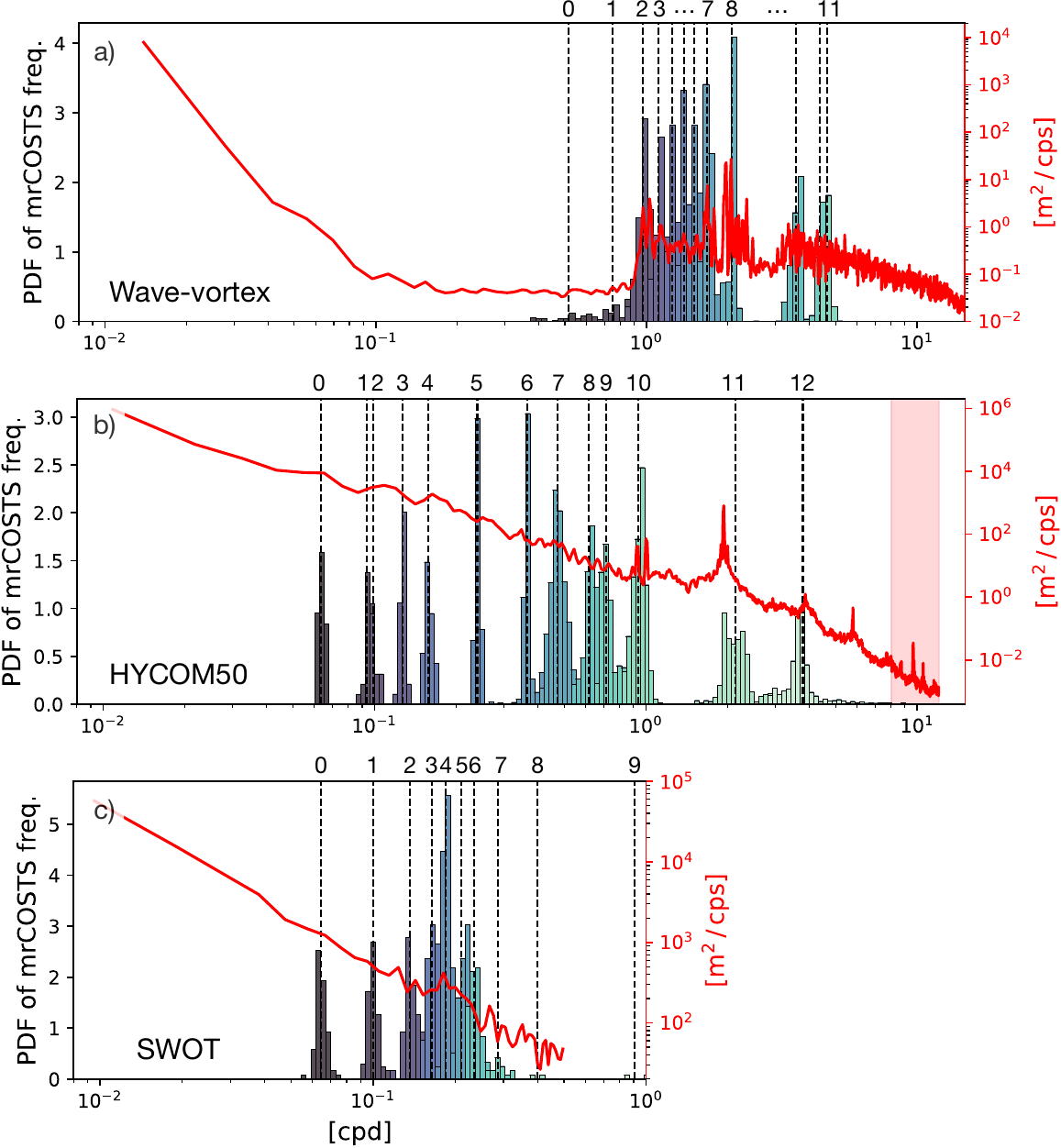}
  \caption{ \label{fig:hist} Probability density function (PDF) of the frequencies in cycles per day associated with mrCOSTS modes, $\text{Im}[\omega^k_{j,\ell}]/2\pi$. Wave-vortex SSHa fields were fed every hour to mrCOSTS and applied in two levels (a). The vertical black dashed lines indicate the frequencies each $k$-means clustering has grouped the mrCOSTS modes around. The top $x$ axis shows the final mrCOSTS bands, viz.~12 bands in total ($p=0,1,...,11$).
  The histogram is colored from light-to-dark shading corresponding to clusters from high-to-low frequency. Frequency spectrum of SSHa in red solid curve is plotted against the right $y$ axis.
  HYCOM50 SSHa snapshot fields were fed every three hours (b). The red shading indicates the three-hour cutoff.
  MrCOSTS discovered 13~bands in total ($p=0,1,...,12$).
  The same but for where instantaneous SWOT data were fed daily and mrCOSTS discovered 10 bands ($p=0,1,...,9$; c). Periodograms were computed every 150 data points along track and 40 data points across track and then spatially averaged in constructing the SWOT frequency spectrum.
  }
\end{figure}

Figure~\ref{fig:hist}b shows the PDF of frequencies discovered by mrCOSTS and the frequency spectrum of SSHa over the three months of August\,-\,October; periodograms were taken every $\sim 100$\,km and then spatially averaged in constructing the spectrum. We see that the SSHa fields contain a strong signal of internal tides with peaks around the diurnal and semidiurnal frequencies.
Focusing on the PDF, one notices that mrCOSTS bands, $\mathcal{G}_p$, around higher-order tidal harmonics are missing. While further tuning of the parameters (e.g.,~number of decomposition levels, window length prescribed to each level, rank of SVD, etc.) may improve the discovery of super-inertial frequencies, given that our interest here is in extracting the sub-inertial geostrophic dynamics, we have settled with the parameter settings described above.

\section{Results}
\subsection{Idealized Wave-Vortex Simulation}
\begin{figure}
    \centering
    \includegraphics[width=0.8\linewidth]{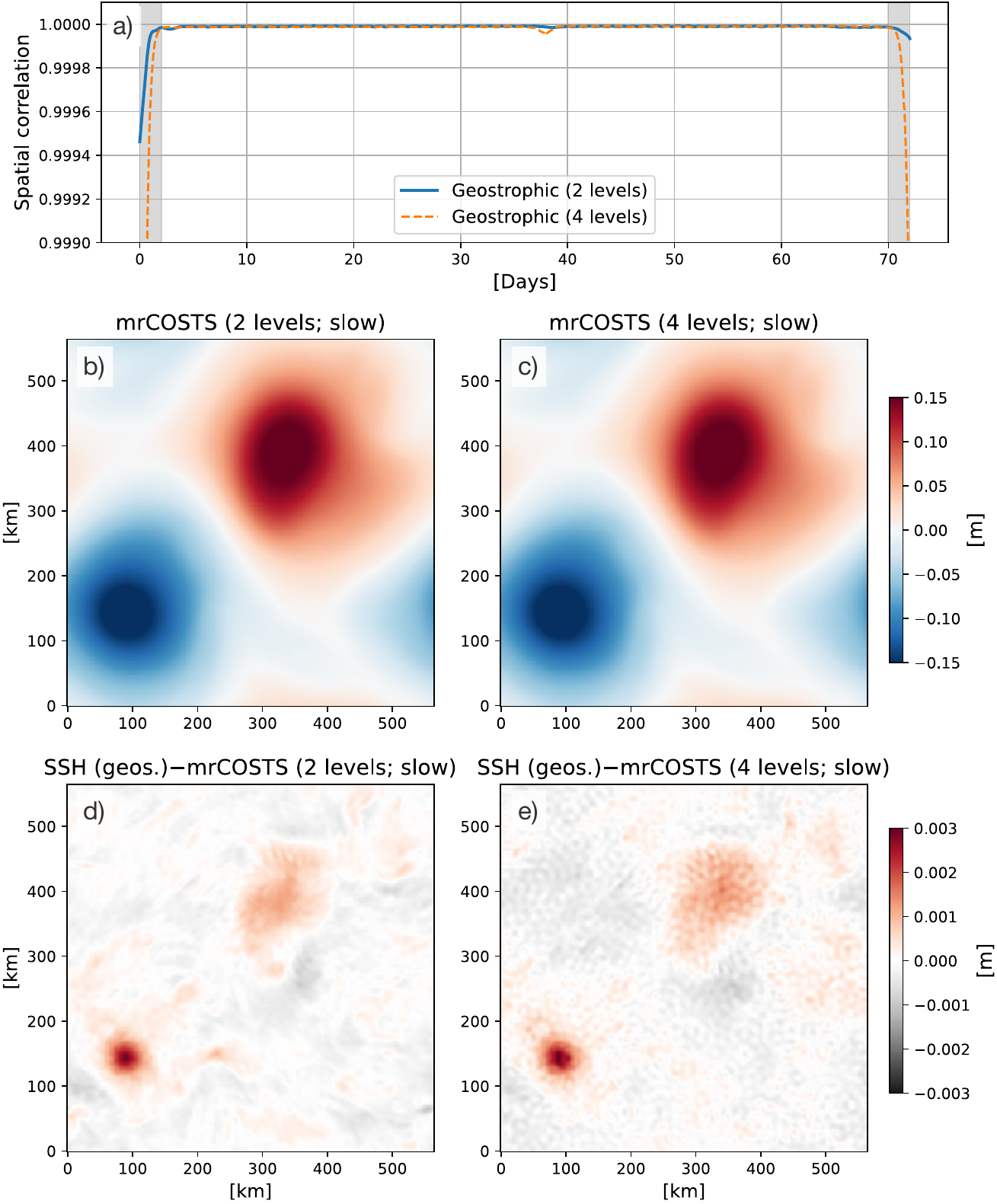}
    \caption{Time series of the spatial correlation between the geostrophic component from the wave-vortex SSH field and mrCOSTS reconstruction of the slow component (a). The black shading indicates the duration of COI.
    A snapshot of the mrCOSTS slow component on the same day as in Fig.~\ref{fig:w-v_SSH}c when mrCOSTS is applied over two levels (b) and four levels (c). 
    The difference between the wave-vortex geostrophic component and mrCOSTS slow component (d,e).}
    \label{fig:w-v_mrCOSTS}
\end{figure}

As a proof of concept, we start by demonstrating the spatial maps of SSHa reconstruction. 
The mrCOSTS reconstruction of the sub-inertial (slow) component of SSHa from the wave-vortex simulation is shown in Fig.~\ref{fig:w-v_mrCOSTS}b.
In total, mrCOSTS discovered 12 bands ($p = 0,1,\cdots,11$) based on the convergence of SVD (Fig.~\ref{fig:hist}a); this is similar to EOF where it yields a finite number of modes ($M$ in \eqref{eq:eof}) or discrete spectral decomposition where the number of Fourier modes is determined by the data resolution and Nyquist frequency/wavenumber.
Given the 12 bands in total, the decision on what to select as part of the `slow' component becomes somewhat subjective.
Here, we chose the slow component as the net sum of the background band and first two bands ($p=0,1$), which are associated with periods longer than a day.
The background (slowest) band is the left-over low-frequency component after finishing recursively applying mrCOSTS as the high-frequency component is extracted at each level.
Figure~\ref{fig:w-v_mrCOSTS}a shows the time series of the spatial correlation between the geostrophic component and mrCOSTS slow component; the correlation is persistently higher than $0.999$.
The difference between Figs.~\ref{fig:w-v_SSH}c and \ref{fig:w-v_mrCOSTS}b,\,c is hardly detectable by the naked eye.
The slightly lower correlation towards the beginning and ending of the time series is attributed to $\breve{\ve}_p(t)$ having the largest errors at the edges of the time domain due to edge effects analogous to the cone-of-influence (COI) in wavelet analysis \cite{torrence1998practical,de2004wavelet,lapo2024multi}.

Now, we can play the game where we assume that we had no prior knowledge of the flow. Namely, a case where, from eye inspection, we can tell that the flow consists of eddies and waves (Fig.~\ref{fig:w-v_SSH}b) but do not know the exact frequencies of the dynamics. Based on the frequency spectrum (Fig.~\ref{fig:hist}a), we can make an educated guess that the flow has peaks about the semidiurnal and diurnal frequencies so we can prescribe the window lengths as [0.5, 1, 2, 16]\,days. The corresponding SVD ranks were set as [4, 8, 10, 18].
MrCOSTS found eight bands in total and we chose the first three bands with periods longer than a day as part of the slow component (Supporting Information Fig.~S1a).
We again find that mrCOSTS decomposes and reconstructs SSHa with small errors (on the order of 1\%; Fig.~\ref{fig:w-v_mrCOSTS}a,\,c and e).
The point of all this is that the mrCOSTS algorithm is highly versatile to the choice of parameters (as was noted in Section~\ref{sec:method}.1) so long as the window lengths cover a reasonable range of distinct time scales for phenomena consisting of multi-scale dynamics \cite{lapo2024multi}.

\subsection{Tidally-Forced Submesoscale-Permitting North Atlantic Simulation}
\begin{figure}[ht]
\centering
 \includegraphics[width=0.85\linewidth]{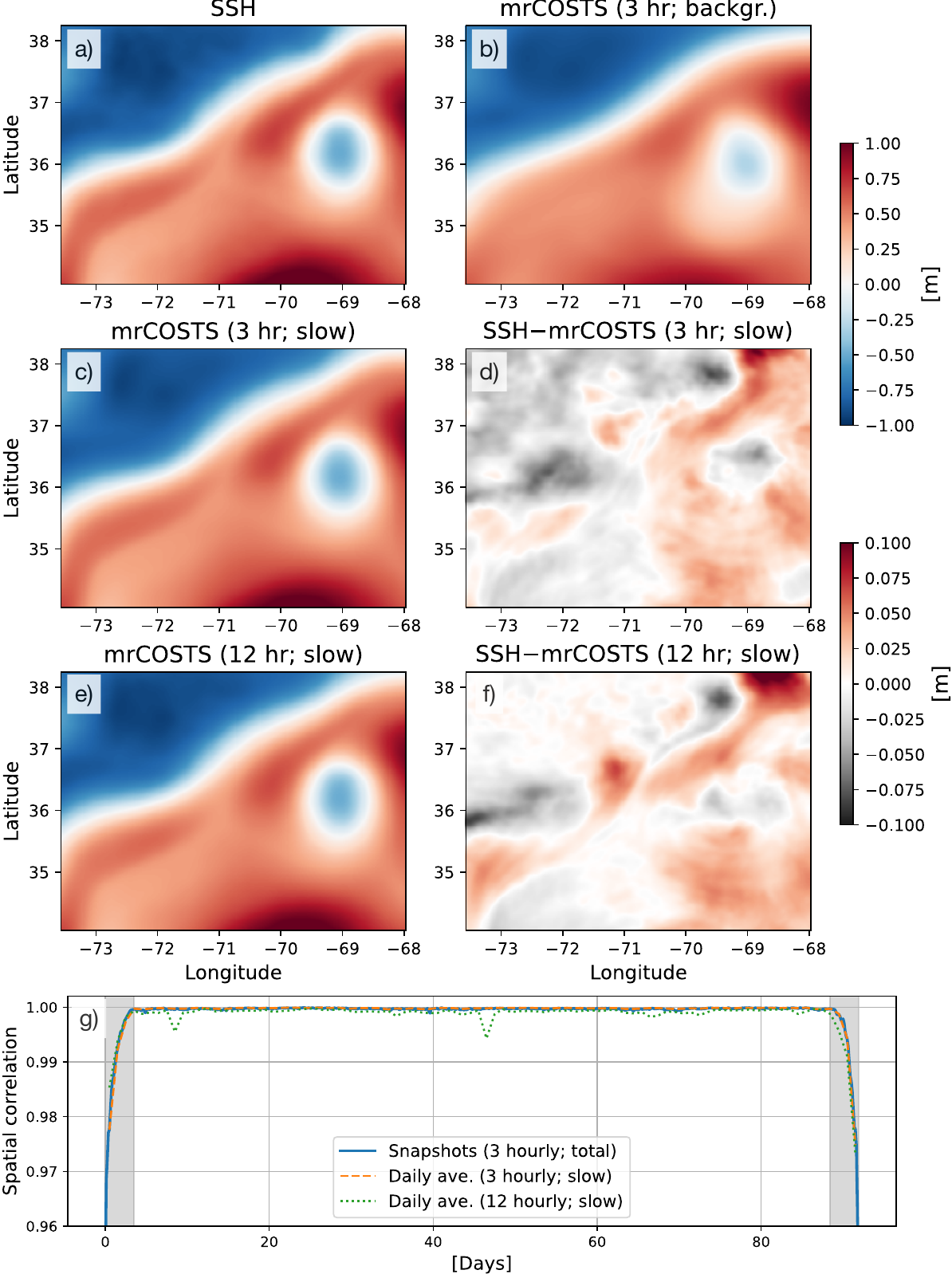}
  \caption{ \label{fig:SSH} Instantaneous snapshots of SSHa and its mrCOSTS reconstruction on an arbitrary day. HYCOM50 output of instantaneous SSHa spatially smoothed using a Gaussian filter with a standard deviation of 10\,km (a), the slowest (background) mrCOSTS band where SSHa fields were fed three hourly (b), mrCOSTS extraction of sub-inertial component (c), residual between HYCOM50 SSHa and sub-inertial component (d). MrCOSTS extraction of the sub-inertial component where SSHa fields were fed 12 hourly (e), and its residual (f).
  Time series of spatial correlation between SSHa and mrCOSTS reconstructions (g). The solid blue curve documents the correlation between instantaneous SSHa and total mrCOSTS reconstruction where SSHa fields were fed three hourly. The orange-dashed and green-dotted curve shows the correlation between daily-averaged SSHa and sub-inertial mrCOSTS reconstruction where data were fed three and 12 hourly respectively.
  The black shading indicates the duration of COI.
  }
\end{figure}

Encouraged by the success from the idealized case, we show the modeled SSHa field in HYCOM50 in Fig.~\ref{fig:SSH}a and the mrCOSTS reconstruction of its sub-inertial (slow) component. 
The slow component was chosen to be the net sum of the background band and bands 0\,-\,7, which have periods longer than 2~days (i.e., frequencies lower than $ 5\times10^{-1}$\,cpd). 
We see that slowest (background) band already captures the large-scale features of the separated Gulf Stream and a cold-core eddy (Fig.~\ref{fig:SSH}b). The addition of bands up to seven further improves the reconstruction when the SSHa fields are fed every three hours to construct $\mat{H}$ and $\mat{H}'$ (Fig.~\ref{fig:SSH}c,\,d). This is corroborated by the spatial correlation shown in Fig.~\ref{fig:SSH}g where outside of COI, the correlation is always higher than $0.99$. 
We also find that the performance of mrCOSTS remains relatively insensitive to temporal sub-sampling. This is highlighted by the spatial maps and spatial correlation where the SSHa fields were given every 12 hours (Fig.~\ref{fig:SSH}e\,-\,g). The first eight bands out of the 10 were summed up to obtain the slow component for the 12-hourly case (Fig.~S1b in Supporting Information).
\label{sec:res}

\begin{figure}[ht]
\centering
 \includegraphics[width=1.\linewidth]{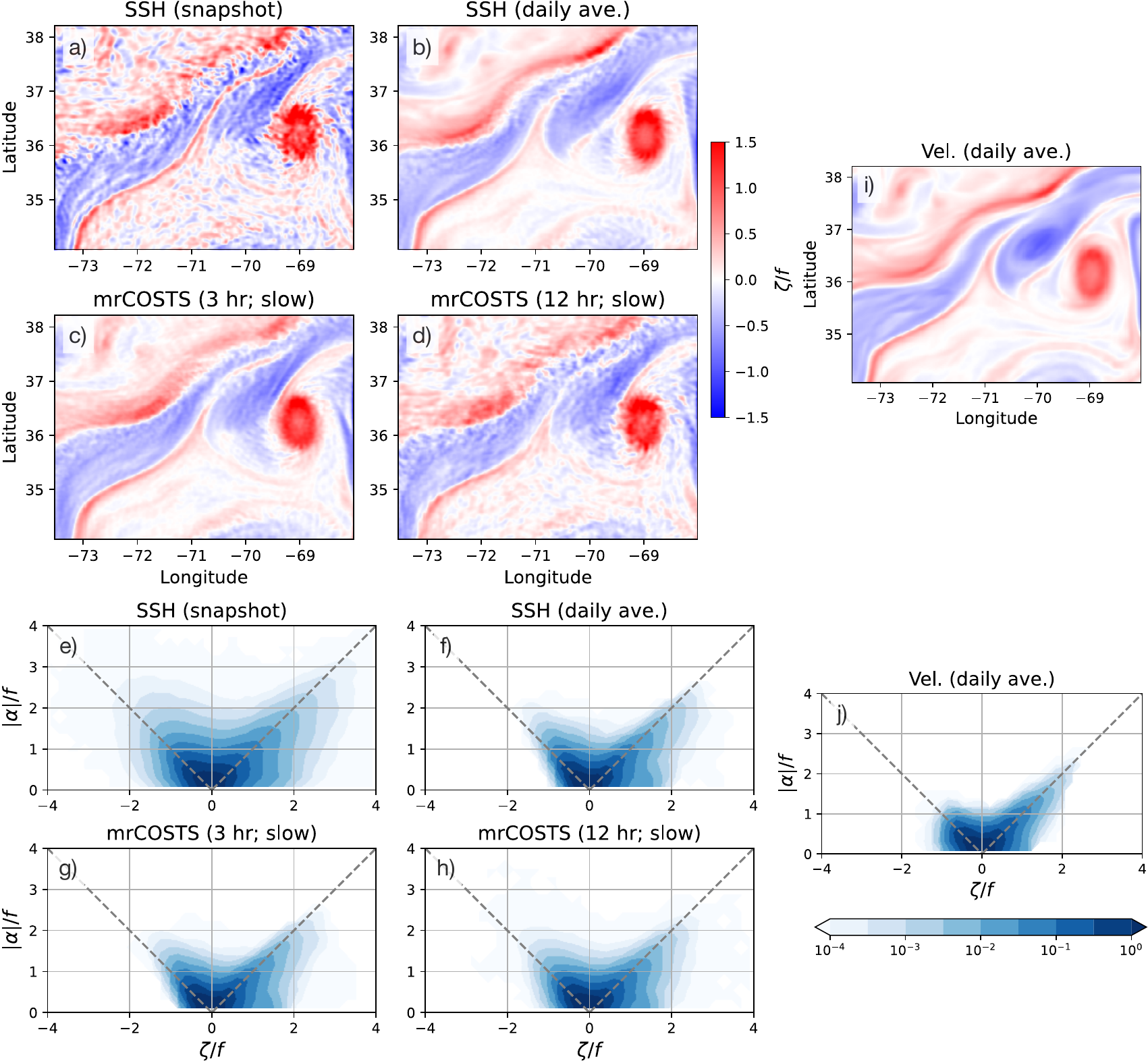}
  \caption{ \label{fig:vort} Spatial maps of relative vorticity normalized by the local Coriolis frequency $\zeta/f$, viz. the local Rossby number Ro from HYCOM50. 
  Panel (a) shows Ro diagnosed from an instantaneous SSHa field spatially smoothed using a Gaussian filter with a standard deviation of 10\,km, and when the hourly SSHa fields are daily averaged (b).
  Instantaneous mrCOSTS reconstructions of the slow component of Ro when data are fed every three and 12 hours are documented in panels (c) and (d).
  Joint probability density functions (PDFs) of Ro and strain rates normalized by $f$ for each case over the three months of August\,--\,October (e\,-\,h).
  A spatial map of Ro and joint PDF of Ro and strain rate at the surface computed from daily-averaged and spatially-smoothed total velocity using a Gaussian filter with a standard deviation of 10\,km is shown for reference (i,\,j).
  }
\end{figure}

Given the extraction of the slow component of SSHa evolution, we can diagnose geostrophy from the fields
\begin{equation}
    \label{eq:geostroph}
    fu = -g\eta_y,\ fv = g\eta_x\,,
\end{equation}
and from it, relative vorticity $\zeta=v_x - u_y$ and strain rate $|\alpha| = \sqrt{(u_x-v_y)^2 + (v_x+u_y)^2}$. Since relative vorticity and strain rates are second-order derivative terms of SSHa, they will highlight the small-scale features \cite<or the lack thereof;>[]{shcherbina2013statistics,balwada2021vertical,jones2023using}.
When the spatially-smoothed instantaneous snapshot outputs of SSHa fields are used, the imprint of internal waves contaminate the estimates of geostrophic relative vorticity (Fig.~\ref{fig:vort}a). This is also indicated in the joint PDF of relative vorticity and strain rates where there is an anomalously high likelihood of values with large amplitude and negative values in relative vorticity (Fig.~\ref{fig:vort}e); geostrophy is only expected to hold under small Rossby numbers \cite{Vallis:2006aa}. The waves can be filtered out by taking daily averages of the hourly SSHa field (Fig.~\ref{fig:vort}a,\,f). This gives us a reference for geostrophic eddies, but we are interested in cases where hourly temporal resolution is not available at hand.

Figure~\ref{fig:vort}c~and~d document the relative vorticity fields from the slow component extracted by mrCOSTS. First thing to note is that the mrCOSTS bands are smooth enough to permit second-order spatial derivatives. There is a large attenuation in the signal from internal waves with its performance being better when SSHa fields are given every three hours compared to 12 hours. The same description also applies to both strain rate and horizontal divergence fields (Figs.~S2~and~S3). Nonetheless, both cases of sub-sampling capture the joint PDF features of geostrophic eddies (Fig.~\ref{fig:vort}g,\,h). The Rossby numbers on the order of unity ($\text{Ro}\sim\mathcal{O}(1)$) present in the mrCOSTS slow component and daily-averaged SSHa likely indicate that the eddies and meandering of the Gulf Stream analysed here are in cyclogeostrophic balance \cite<Fig.~\ref{fig:vort}b,\,c;>[]{Hironetal_2021}, which are signals we want to retain in addition to geostrophy.

\section{Discussion and Conclusions}
\label{sec:conc}
We end by discussing results on applying multi-resolution coherent spatiotemporal scale separation (mrCOSTS) to the one-day-repeat SWOT observations of SSHa ($\Delta t = 24$~hours) during its Calibration and Validation (Cal/Val) phase. We have taken the Level 3 (L3) KaRIn filtered product as our interest here is in extracting the first-order balance, geostrophy, from signals that include internal tides. The domain we use is between $30^\circ$\,-\,$40^\circ$N and $284^\circ$\,-\,$288^\circ$E for pass number nine situated across the separated Gulf Stream path. Missing data and spacing between the swaths were linearly interpolated over and when data were missing from over 70\% of the swaths, that day was dropped and temporally interpolated over between the day before and after via a linear spline. The SWOT SSHa fields were further smoothed with a Gaussian filter with the standard deviation of 15\,km. MrCOSTS was then applied with the period associated with each window length prescribed as $[9,10,11,30]$\,days and rank of SVD as $[4,4,6,10]$ respectively.
The slow component was defined as the sum of the background and first six bands (0\,-\,5) out of the 10 (Fig.~\ref{fig:hist}c).

\begin{figure}[ht]
    \centering
    \includegraphics[width=0.9\linewidth]{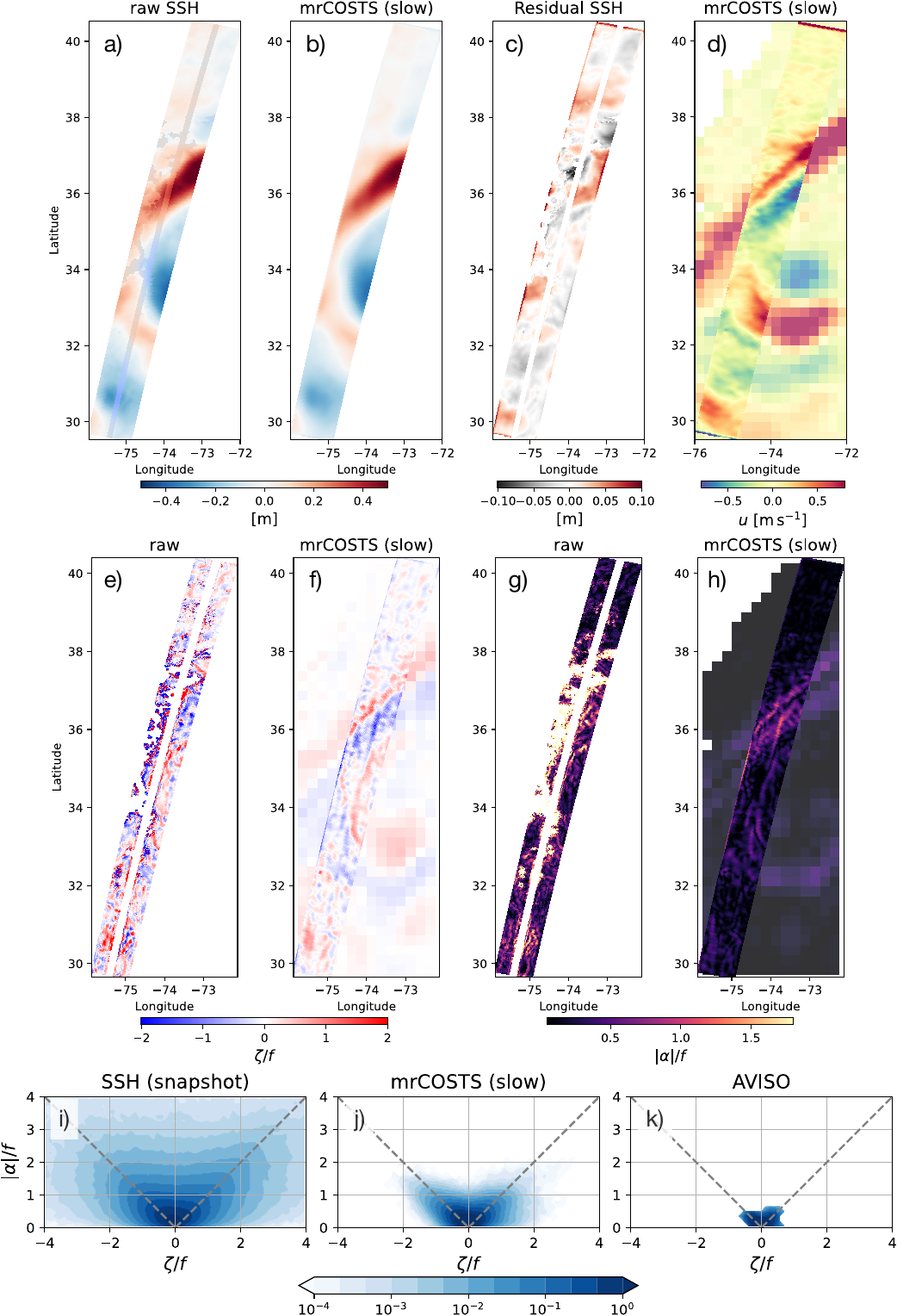}
    \caption{ \label{fig:swot} L3 SWOT observation of SSHa on June 21, 2023 (a), mrCOSTS reconstruction of the slow component of the spatially filtered SSHa (b), and the difference between the two (c). Missing data and spacing between the swaths are interpolated over for SWOT SSHa and shown in a different colormap in panel (a).
    Zonal geostrophic velocity diagnosed from the mrCOSTS slow component (d).
    Relative vorticity $\zeta$ and strain rate $|\alpha|$ normalized by $f$ diagnosed from the SWOT data and mrCOSTS slow component (e\,-\,f).
    Geostrophic zonal velocity, relative vorticity and strain rate from daily-averaged $0.25^\circ$~gridded AVISO are shown in lighter shadings in contrast to mrCOSTS.
    Joint PDF of $\zeta/f$ and $|\alpha|/f$ diagnosed from raw SWOT data, mrCOSTS slow component and AVISO (i\,-\,k).
    The SWOT fields used in panels (a,e,g,i) were not spatially filtered as the L3 product \textit{a priori} has some smoothing applied \cite{dibarboure2024blend}.
  }
\end{figure}

When geostrophy \eqref{eq:geostroph} is applied directly to SWOT data, the relative vorticity and strain rate take fictitiously large values in magnitude despite the L3 product being somewhat smoothed via the de-noising process \cite<Fig.~\ref{fig:swot}e,\,g,\,i;>[]{dibarboure2024blend}; in hindsight, the large magnitudes may have been expected as we are applying \eqref{eq:geostroph} to a field that includes signals of super-inertial balanced and unbalanced dynamics.
The mrCOSTS slow component of SSHa, on the other hand, captures the large-scale feature and is much smoother than the SWOT data (Fig.~\ref{fig:swot}a,\,b). The zonal geostrophic velocity from the mrCOSTS slow component captures the separated Gulf Stream about 37$^\circ$N (Fig.~\ref{fig:swot}c), and the fields of relative vorticity and strain rate become smoother and fall within the acceptable range of magnitude (Fig.~\ref{fig:swot}f,\,h).
Nonetheless, the joint PDF does not capture the skewness towards positive relative vorticity values (Fig.~\ref{fig:swot}j).
As a reference, the joint PDF computed from daily-averaged $0.25^\circ$~gridded AVISO data during April\,--\,June, 2023 is shown in Fig.~\ref{fig:swot}k, which does not capture the skewness either.
The spatial correlation between SWOT SSHa and its mrCOSTS reconstruction is generally higher than 0.9 during the Cal/Val phase (Fig.~S4) but is worse than the case with the wave-vortex and HYCOM50 simulations.
While there is a hint of mrCOSTS detecting the diurnal tidal signal (band nine in Fig.~\ref{fig:hist}c), it is likely that the duration of three months with daily resolution (i.e.,~$102$~data points in time) is not a sufficient amount of data to robustly estimate $\mat{A}$ from the least-squares fit \eqref{eq:A}.

\begin{figure}
    \centering
    \includegraphics[width=0.8\linewidth]{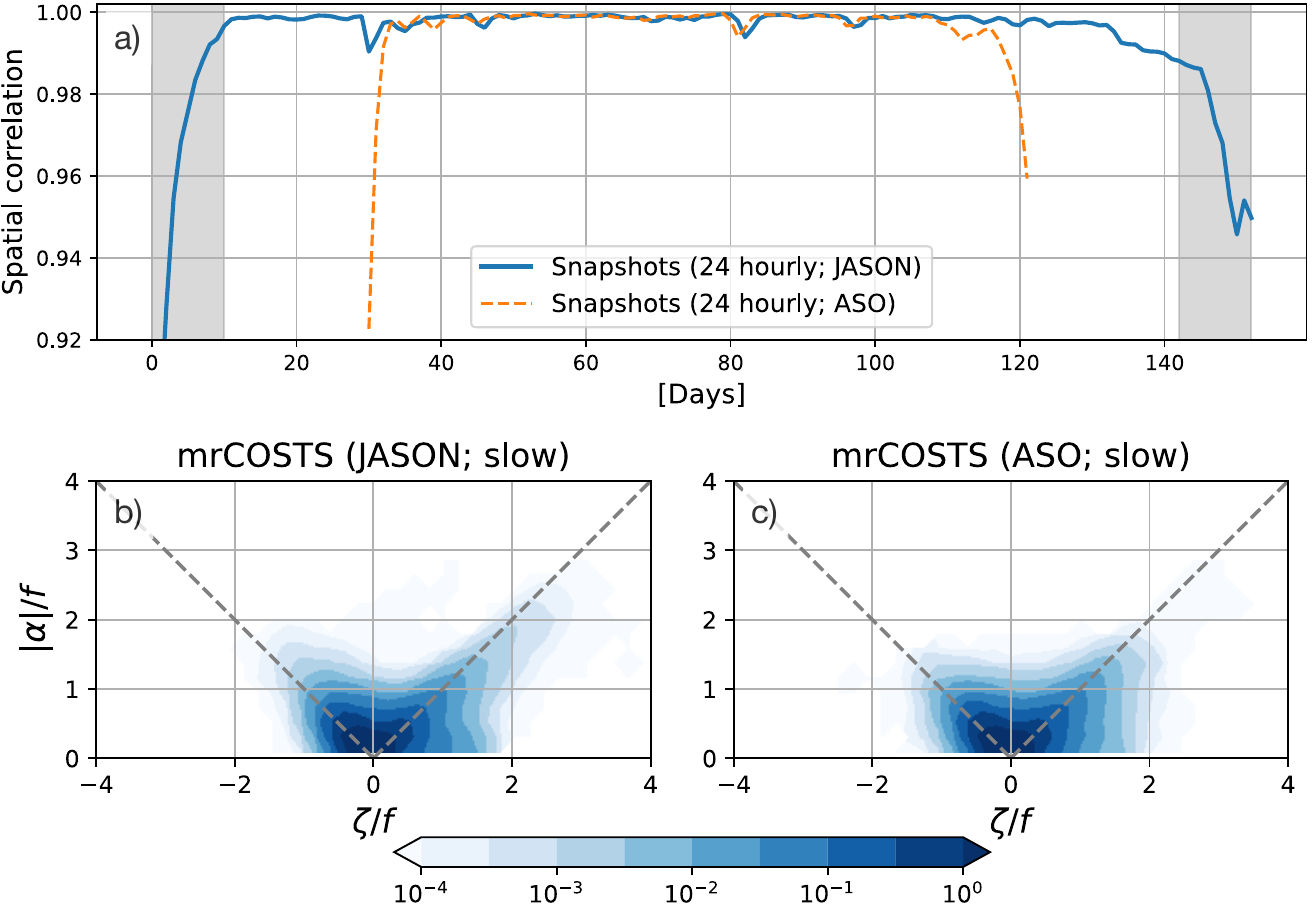}
    \caption{Time series of spatial correlation between HYCOM50 SSHa and its total reconstruction by mrCOSTS when data is fed 24 hourly (a). The black shading indicates the duration of COI for the JASON case. Joint PDFs of relative vorticity and strain rate normalized by the local Coriolis frequency for the JASON (b) and ASO case (c).}
    \label{fig:HY24}
\end{figure}

In order to test whether extending the duration of the data would improve the extraction of geostrophy, we examine the mrCOSTS reconstruction of HYCOM50 SSHa snapshot fields taken at daily intervals ($\Delta t = 24$\,hours) when the duration to construct $\mat{H}$ and $\mat{H}'$ is taken over the three months of August to October (ASO), and five months of July to November (JASON).
The window lengths were prescribed as $[9, 10, 11, 30, 90]$\,days and SVD ranks as $[4, 4, 6, 10, 18]$ for the JASON case where 90~days corresponds to seasonal time scales. MrCOSTS was applied over four levels using the first four parameters for the ASO case.
MrCOSTS discovered seven bands in total for the JASON case ($p=0,1,\cdots,6$; Fig.~S1c) and six bands for the ASO case ($p=0,1,\cdots,5$; Fig.~S1d).
We find that in both cases, mrCOSTS is able to reconstruct the HYCOM50 SSHa fields relatively well; the spatial correlation outside of COI is higher than 0.98 (Fig.~\ref{fig:HY24}a).
For the slow component, the sum of the first five bands were chosen for JASON and of the first four bands for ASO in addition to their respective background bands.
We again diagnose the geostrophic relative vorticity and strain rates from the slow components and the joint PDFs of the two are documented in Fig.~\ref{fig:HY24}b,\,c.
Similar to SWOT (Fig.~\ref{fig:swot}j), the joint PDF from the ASO case does not present the skewness in relative vorticity.
However, we find that the JASON case is able to recover the skewness and and the joint PDF becomes closer to Fig.~\ref{fig:vort}f.
This corroborates our hypothesis that the performance of mrCOSTS depends somewhat on the number data points in time to fit \eqref{eq:A}.
This sensitivity to the volume and quality of data is not unique to DMD but rather universal to data-driven methods \cite<e.g.>[]{chen2023machine,smith2023temporal,mojgani2024interpretable}.

The goals of this paper were to introduce mrCOSTS, a variant of dynamic-mode decomposition (DMD), to the oceanographic and earth science community.
While machine-learning methods have shown some promise in extracting the surface flow kinematics from SSH \cite<e.g.,>[]{sinha2021estimating,xiao2023reconstruction,gao2024deep,archambault2024learning,cutolo2024cloinet,fablet2024inversion,febvre2024training,martin2024deep,lyu2024multi}, we have opted for DMD here due its interpretability owing to it essentially being a combination of linear-algebraic operations. The fact that DMD naturally decomposes the data into frequency components is also well suited for disentangling geostrophically balanced motions from internal waves where the two tend to have distinct characteristic time scales.
We have showcased that by applying mrCOSTS to modeled and observed SSHa, its slow bands are usable to diagnose geostrophy.
In contrast to other DMD-based methods, mrCOSTS is able to robustly extract spatially coherent spatial modes, which are smooth enough to permit spatial derivatives \cite{lapo2024multi}.
The need for scale-separation methods is wide spread in the general earth science community; for example, it would be interesting to apply mrCOSTS to long-standing issues such as quantifying orographic precipitation patterns \cite<e.g.,>[]{buttafuoco2011spatial,curio2016seasonality,li2024snow} or discovering climate modes \cite<e.g.,>[]{newman2016pacific,dewar2022routine,mishonov2024revisiting,wang2024indian,miyamoto2024low} and eddy parametrizations \cite{li2023stochastic}.

Future work involves extending our analyses to other geographical regions, the 21-day-repeat SWOT orbit where data will be available for longer periods than three months, and to extract higher-order balances than geostrophy, viz.,~quasi- and semi-geostrophy.
The Southern Ocean may be an appealing region given the overlap amongst SWOT swaths increases compared to lower latitudes.
While we have purely focused on the geostrophically balanced component of the flow in this study, it is true that information on the unbalanced motion (e.g.,~internal waves) is also of significant value \cite{yadidya2024phase,dematteis2024interacting,tchilibou2024internal}. 
It is unclear to what extent DMDs can separate out internal waves from submesoscale dynamics \cite<or waves from turbulence in general; cf.>[]{chavez2024wave}, which tend to be associated with similar time scales, but will be an avenue for further investigation.

\section*{Open Research}
The wave-vortex model is available from \citeA<>[\url{https://github.com/Energy-Pathways-Group/GLOceanKit}]{early2024wvm} and the scripts for this particular simulation are available at \url{https://github.com/JeffreyEarly/DMDEddySimulation}. 
The HYCOM50 model outputs used in this paper are publicly available on the Open Storage Network (OSN; \url{https://www.openstoragenetwork.org/}). Example Jupyter notebooks and Yaml file used to access the data are available on Github \cite<>[\url{https://github.com/pangeo-data/swot_adac_ogcms}]{swotadacogcm2022}.
Jupyter notebooks used for analyses will be shared with a DOI upon acceptance of the manuscript (\url{https://github.com/roxyboy/GeostrophicDMD/tree/div}).
Level 3 SWOT data were accessed from \citeA<>[\url{https://www.aviso.altimetry.fr/en/data/products/sea-surface-height-products/global/swot-l3-ocean-products.html}]{dibarboure2024blend}.






\acknowledgments
We graciously thank the developers of {\tt PyDMD}, an open-source Python package used to apply DMDs \cite<>[\url{https://github.com/PyDMD/PyDMD}]{demo2018pydmd,ichinaga2024pydmd}. 
Fourier spectra were computed using the {\tt xrft} and joint PDFs using the {\tt xhistogram} Python packages respectively \cite{xrft2021,xhist2021}.
Spatial filtering was taken using the {\tt gcm-filters} Python package \cite{grooms2021diffusion,loose2022gcm}.
We thank the COAPS modeling group for making their model outputs publicly available via OSN.
The altimeter products were produced by Ssalto/Duacs and distributed by $\text{AVISO}+$, with support from CNES (\url{https://www.aviso.altimetry.fr}).
This work is a contribution to the SWOT Science Team. T.~Uchida's doctoral education back in 2014\,--\,2019 (before the world disfigured due to Covid) was primarily funded by the SWOT-related NASA award NNX16AJ35G; it is interesting to reminisce on the fact that he has come full circle to work with SWOT again. 
Uchida acknowledges support from the NSF grants OCE-2123632 and OCE-1941963.
B.\,K.~Arbic acknowledges support from the NASA grant 80NSSC20K1135. 
J.\,J.~Early’s work was supported by the NSF grant OCE-2123740 and NASA award 80NSSC21K1193.
B.~Yadidya acknowledges funding support from Office of Naval Research (ONR) grant N00017-22-1-2576.
We would like to extend our gratitude to Edward Peirce and Kelly Hirai for maintaining the Florida State University cluster on which the data were analyzed.


%
\bibliography{Eddy-param} 

\begin{thebibliography}{}

\bibitem [\protect \citeauthoryear {%
Abernathey%
\ \protect \BOthers {.}}{%
Abernathey%
\ \protect \BOthers {.}}{%
{\protect \APACyear {2023}}%
}]{%
xhist2021}
\APACinsertmetastar {%
xhist2021}%
\begin{APACrefauthors}%
Abernathey, R\BPBI P.%
, Squire, D.%
, Bourbeau, J.%
, Nicholas, T.%
, Bourbeau, J.%
, Joseph, G.%
\BDBL {}others%
\end{APACrefauthors}%
\unskip\
\newblock
\APACrefYearMonthDay{2023}{}{}.
\newblock
\APACrefbtitle {{\tt xhistogram}: Fast, flexible, label-aware histograms for numpy and xarray [{S}oftware].} {{\tt xhistogram}: Fast, flexible, label-aware histograms for numpy and xarray [{S}oftware].}
\newblock
\APAChowpublished {Zenodo}.
\newblock
\begin{APACrefURL} \url{https://xhistogram.readthedocs.io/en/latest/} \end{APACrefURL}
\newblock
\begin{APACrefDOI} \doi{10.5281/zenodo.7095156} \end{APACrefDOI}
\PrintBackRefs{\CurrentBib}

\bibitem [\protect \citeauthoryear {%
Archambault%
, Filoche%
, Charantonis%
, B{\'e}r{\'e}ziat%
\BCBL {}\ \BBA {} Thiria%
}{%
Archambault%
\ \protect \BOthers {.}}{%
{\protect \APACyear {2024}}%
}]{%
archambault2024learning}
\APACinsertmetastar {%
archambault2024learning}%
\begin{APACrefauthors}%
Archambault, T.%
, Filoche, A.%
, Charantonis, A.%
, B{\'e}r{\'e}ziat, D.%
\BCBL {}\ \BBA {} Thiria, S.%
\end{APACrefauthors}%
\unskip\
\newblock
\APACrefYearMonthDay{2024}{}{}.
\newblock
{\BBOQ}\APACrefatitle {Learning sea surface height interpolation from multi-variate simulated satellite observations} {Learning sea surface height interpolation from multi-variate simulated satellite observations}.{\BBCQ}
\newblock
\APACjournalVolNumPages{Journal of Advances in Modeling Earth Systems}{16}{6}{e2023MS004047}.
\newblock
\begin{APACrefDOI} \doi{10.1029/2023MS004047} \end{APACrefDOI}
\PrintBackRefs{\CurrentBib}

\bibitem [\protect \citeauthoryear {%
Askham%
\ \BBA {} Kutz%
}{%
Askham%
\ \BBA {} Kutz%
}{%
{\protect \APACyear {2018}}%
}]{%
askham2018variable}
\APACinsertmetastar {%
askham2018variable}%
\begin{APACrefauthors}%
Askham, T.%
\BCBT {}\ \BBA {} Kutz, J\BPBI N.%
\end{APACrefauthors}%
\unskip\
\newblock
\APACrefYearMonthDay{2018}{}{}.
\newblock
{\BBOQ}\APACrefatitle {Variable projection methods for an optimized dynamic mode decomposition} {Variable projection methods for an optimized dynamic mode decomposition}.{\BBCQ}
\newblock
\APACjournalVolNumPages{SIAM Journal on Applied Dynamical Systems}{17}{1}{380--416}.
\newblock
\begin{APACrefDOI} \doi{10.1137/M1124176} \end{APACrefDOI}
\PrintBackRefs{\CurrentBib}

\bibitem [\protect \citeauthoryear {%
Baddoo%
, Herrmann%
, McKeon%
, Nathan~Kutz%
\BCBL {}\ \BBA {} Brunton%
}{%
Baddoo%
\ \protect \BOthers {.}}{%
{\protect \APACyear {2023}}%
}]{%
baddoo2023physics}
\APACinsertmetastar {%
baddoo2023physics}%
\begin{APACrefauthors}%
Baddoo, P\BPBI J.%
, Herrmann, B.%
, McKeon, B\BPBI J.%
, Nathan~Kutz, J.%
\BCBL {}\ \BBA {} Brunton, S\BPBI L.%
\end{APACrefauthors}%
\unskip\
\newblock
\APACrefYearMonthDay{2023}{}{}.
\newblock
{\BBOQ}\APACrefatitle {Physics-informed dynamical mode decomposition} {Physics-informed dynamical mode decomposition}.{\BBCQ}
\newblock
\APACjournalVolNumPages{Proceedings of the Royal Society A}{479}{2271}{20220576}.
\newblock
\begin{APACrefDOI} \doi{10.1098/rspa.2022.0576} \end{APACrefDOI}
\PrintBackRefs{\CurrentBib}

\bibitem [\protect \citeauthoryear {%
Baker%
, Kafiabad%
\BCBL {}\ \BBA {} Vanneste%
}{%
Baker%
\ \protect \BOthers {.}}{%
{\protect \APACyear {2024}}%
}]{%
baker2024lagrangian}
\APACinsertmetastar {%
baker2024lagrangian}%
\begin{APACrefauthors}%
Baker, L\BPBI E.%
, Kafiabad, H\BPBI A.%
\BCBL {}\ \BBA {} Vanneste, J.%
\end{APACrefauthors}%
\unskip\
\newblock
\APACrefYearMonthDay{2024}{}{}.
\newblock
{\BBOQ}\APACrefatitle {Lagrangian filtering for wave-mean flow decomposition} {Lagrangian filtering for wave-mean flow decomposition}.{\BBCQ}
\newblock
\APACjournalVolNumPages{arXiv preprint arXiv:2406.03477}{}{}{}.
\PrintBackRefs{\CurrentBib}

\bibitem [\protect \citeauthoryear {%
Bakhoday~Paskyabi%
}{%
Bakhoday~Paskyabi%
}{%
{\protect \APACyear {2024}}%
}]{%
bakhoday2024impact}
\APACinsertmetastar {%
bakhoday2024impact}%
\begin{APACrefauthors}%
Bakhoday~Paskyabi, M.%
\end{APACrefauthors}%
\unskip\
\newblock
\APACrefYearMonthDay{2024}{}{}.
\newblock
{\BBOQ}\APACrefatitle {Impact of swell waves on atmospheric surface turbulence: wave--turbulence decomposition methods} {Impact of swell waves on atmospheric surface turbulence: wave--turbulence decomposition methods}.{\BBCQ}
\newblock
\APACjournalVolNumPages{Wind Energy Science}{9}{8}{1631--1645}.
\newblock
\begin{APACrefDOI} \doi{10.5194/wes-9-1631-2024} \end{APACrefDOI}
\PrintBackRefs{\CurrentBib}

\bibitem [\protect \citeauthoryear {%
Balwada%
, Xiao%
, Smith%
, Abernathey%
\BCBL {}\ \BBA {} Gray%
}{%
Balwada%
\ \protect \BOthers {.}}{%
{\protect \APACyear {2021}}%
}]{%
balwada2021vertical}
\APACinsertmetastar {%
balwada2021vertical}%
\begin{APACrefauthors}%
Balwada, D.%
, Xiao, Q.%
, Smith, K\BPBI S.%
, Abernathey, R\BPBI P.%
\BCBL {}\ \BBA {} Gray, A\BPBI R.%
\end{APACrefauthors}%
\unskip\
\newblock
\APACrefYearMonthDay{2021}{}{}.
\newblock
{\BBOQ}\APACrefatitle {Vertical fluxes conditioned on vorticity and strain reveal submesoscale ventilation} {Vertical fluxes conditioned on vorticity and strain reveal submesoscale ventilation}.{\BBCQ}
\newblock
\APACjournalVolNumPages{Journal of Physical Oceanography}{51}{9}{2883--2901}.
\newblock
\begin{APACrefDOI} \doi{10.1175/JPO-D-21-0016.1} \end{APACrefDOI}
\PrintBackRefs{\CurrentBib}

\bibitem [\protect \citeauthoryear {%
B\BPBI W.~Brunton%
, Johnson%
, Ojemann%
\BCBL {}\ \BBA {} Kutz%
}{%
B\BPBI W.~Brunton%
\ \protect \BOthers {.}}{%
{\protect \APACyear {2016}}%
}]{%
brunton2016extracting}
\APACinsertmetastar {%
brunton2016extracting}%
\begin{APACrefauthors}%
Brunton, B\BPBI W.%
, Johnson, L\BPBI A.%
, Ojemann, J\BPBI G.%
\BCBL {}\ \BBA {} Kutz, J\BPBI N.%
\end{APACrefauthors}%
\unskip\
\newblock
\APACrefYearMonthDay{2016}{}{}.
\newblock
{\BBOQ}\APACrefatitle {Extracting spatial--temporal coherent patterns in large-scale neural recordings using dynamic mode decomposition} {Extracting spatial--temporal coherent patterns in large-scale neural recordings using dynamic mode decomposition}.{\BBCQ}
\newblock
\APACjournalVolNumPages{Journal of neuroscience methods}{258}{}{1--15}.
\newblock
\begin{APACrefDOI} \doi{10.1016/j.jneumeth.2015.10.010} \end{APACrefDOI}
\PrintBackRefs{\CurrentBib}

\bibitem [\protect \citeauthoryear {%
S\BPBI L.~Brunton%
\ \BBA {} Kutz%
}{%
S\BPBI L.~Brunton%
\ \BBA {} Kutz%
}{%
{\protect \APACyear {2024}}%
}]{%
brunton2024promising}
\APACinsertmetastar {%
brunton2024promising}%
\begin{APACrefauthors}%
Brunton, S\BPBI L.%
\BCBT {}\ \BBA {} Kutz, J\BPBI N.%
\end{APACrefauthors}%
\unskip\
\newblock
\APACrefYearMonthDay{2024}{}{}.
\newblock
{\BBOQ}\APACrefatitle {Promising directions of machine learning for partial differential equations} {Promising directions of machine learning for partial differential equations}.{\BBCQ}
\newblock
\APACjournalVolNumPages{Nature Computational Science}{}{}{1--12}.
\newblock
\begin{APACrefDOI} \doi{10.1038/s43588-024-00643-2} \end{APACrefDOI}
\PrintBackRefs{\CurrentBib}

\bibitem [\protect \citeauthoryear {%
Buttafuoco%
, Caloiero%
\BCBL {}\ \BBA {} Coscarelli%
}{%
Buttafuoco%
\ \protect \BOthers {.}}{%
{\protect \APACyear {2011}}%
}]{%
buttafuoco2011spatial}
\APACinsertmetastar {%
buttafuoco2011spatial}%
\begin{APACrefauthors}%
Buttafuoco, G.%
, Caloiero, T.%
\BCBL {}\ \BBA {} Coscarelli, R.%
\end{APACrefauthors}%
\unskip\
\newblock
\APACrefYearMonthDay{2011}{}{}.
\newblock
{\BBOQ}\APACrefatitle {{Spatial and temporal patterns of the mean annual precipitation at decadal time scale in southern Italy (Calabria region)}} {{Spatial and temporal patterns of the mean annual precipitation at decadal time scale in southern Italy (Calabria region)}}.{\BBCQ}
\newblock
\APACjournalVolNumPages{Theoretical and Applied Climatology}{105}{}{431--444}.
\newblock
\begin{APACrefDOI} \doi{10.1007/s00704-011-0398-8} \end{APACrefDOI}
\PrintBackRefs{\CurrentBib}

\bibitem [\protect \citeauthoryear {%
Cao%
, Fox-Kemper%
\BCBL {}\ \BBA {} Jing%
}{%
Cao%
\ \protect \BOthers {.}}{%
{\protect \APACyear {2021}}%
}]{%
cao2021submesoscale}
\APACinsertmetastar {%
cao2021submesoscale}%
\begin{APACrefauthors}%
Cao, H.%
, Fox-Kemper, B.%
\BCBL {}\ \BBA {} Jing, Z.%
\end{APACrefauthors}%
\unskip\
\newblock
\APACrefYearMonthDay{2021}{}{}.
\newblock
{\BBOQ}\APACrefatitle {{Submesoscale eddies in the upper ocean of the Kuroshio Extension from high-resolution simulation: Energy budget}} {{Submesoscale eddies in the upper ocean of the Kuroshio Extension from high-resolution simulation: Energy budget}}.{\BBCQ}
\newblock
\APACjournalVolNumPages{Journal of Physical Oceanography}{51}{7}{2181--2201}.
\newblock
\begin{APACrefDOI} \doi{10.1175/JPO-D-20-0267.1} \end{APACrefDOI}
\PrintBackRefs{\CurrentBib}

\bibitem [\protect \citeauthoryear {%
Ch{\'a}vez-Dorado%
, Scherl%
\BCBL {}\ \BBA {} DiBenedetto%
}{%
Ch{\'a}vez-Dorado%
\ \protect \BOthers {.}}{%
{\protect \APACyear {2024}}%
}]{%
chavez2024wave}
\APACinsertmetastar {%
chavez2024wave}%
\begin{APACrefauthors}%
Ch{\'a}vez-Dorado, J.%
, Scherl, I.%
\BCBL {}\ \BBA {} DiBenedetto, M.%
\end{APACrefauthors}%
\unskip\
\newblock
\APACrefYearMonthDay{2024}{}{}.
\newblock
{\BBOQ}\APACrefatitle {Wave and turbulence separation using dynamic mode decomposition} {Wave and turbulence separation using dynamic mode decomposition}.{\BBCQ}
\newblock
\APACjournalVolNumPages{arXiv preprint arXiv:2403.00223}{}{}{}.
\newblock
\begin{APACrefDOI} \doi{10.48550/arXiv.2403.00223} \end{APACrefDOI}
\PrintBackRefs{\CurrentBib}

\bibitem [\protect \citeauthoryear {%
Chen%
\ \protect \BOthers {.}}{%
Chen%
\ \protect \BOthers {.}}{%
{\protect \APACyear {2023}}%
}]{%
chen2023machine}
\APACinsertmetastar {%
chen2023machine}%
\begin{APACrefauthors}%
Chen, L.%
, Han, B.%
, Wang, X.%
, Zhao, J.%
, Yang, W.%
\BCBL {}\ \BBA {} Yang, Z.%
\end{APACrefauthors}%
\unskip\
\newblock
\APACrefYearMonthDay{2023}{}{}.
\newblock
{\BBOQ}\APACrefatitle {Machine learning methods in weather and climate applications: A survey} {Machine learning methods in weather and climate applications: A survey}.{\BBCQ}
\newblock
\APACjournalVolNumPages{Applied Sciences}{13}{21}{12019}.
\newblock
\begin{APACrefDOI} \doi{10.3390/app132112019} \end{APACrefDOI}
\PrintBackRefs{\CurrentBib}

\bibitem [\protect \citeauthoryear {%
Chi-Dur{\'a}n%
\ \BBA {} Buffett%
}{%
Chi-Dur{\'a}n%
\ \BBA {} Buffett%
}{%
{\protect \APACyear {2023}}%
}]{%
chi2023extracting}
\APACinsertmetastar {%
chi2023extracting}%
\begin{APACrefauthors}%
Chi-Dur{\'a}n, R.%
\BCBT {}\ \BBA {} Buffett, B\BPBI A.%
\end{APACrefauthors}%
\unskip\
\newblock
\APACrefYearMonthDay{2023}{}{}.
\newblock
{\BBOQ}\APACrefatitle {Extracting spatial--temporal coherent patterns in geomagnetic secular variation using dynamic mode decomposition} {Extracting spatial--temporal coherent patterns in geomagnetic secular variation using dynamic mode decomposition}.{\BBCQ}
\newblock
\APACjournalVolNumPages{Geophysical Research Letters}{50}{5}{e2022GL101288}.
\newblock
\begin{APACrefDOI} \doi{10.1029/2022GL101288} \end{APACrefDOI}
\PrintBackRefs{\CurrentBib}

\bibitem [\protect \citeauthoryear {%
Curio%
\ \BBA {} Scherer%
}{%
Curio%
\ \BBA {} Scherer%
}{%
{\protect \APACyear {2016}}%
}]{%
curio2016seasonality}
\APACinsertmetastar {%
curio2016seasonality}%
\begin{APACrefauthors}%
Curio, J.%
\BCBT {}\ \BBA {} Scherer, D.%
\end{APACrefauthors}%
\unskip\
\newblock
\APACrefYearMonthDay{2016}{}{}.
\newblock
{\BBOQ}\APACrefatitle {{Seasonality and spatial variability of dynamic precipitation controls on the Tibetan Plateau}} {{Seasonality and spatial variability of dynamic precipitation controls on the Tibetan Plateau}}.{\BBCQ}
\newblock
\APACjournalVolNumPages{Earth System Dynamics}{7}{3}{767--782}.
\newblock
\begin{APACrefDOI} \doi{10.5194/esd-7-767-2016} \end{APACrefDOI}
\PrintBackRefs{\CurrentBib}

\bibitem [\protect \citeauthoryear {%
Cutolo%
, Pascual%
, Ruiz%
, Zarokanellos%
\BCBL {}\ \BBA {} Fablet%
}{%
Cutolo%
\ \protect \BOthers {.}}{%
{\protect \APACyear {2024}}%
}]{%
cutolo2024cloinet}
\APACinsertmetastar {%
cutolo2024cloinet}%
\begin{APACrefauthors}%
Cutolo, E.%
, Pascual, A.%
, Ruiz, S.%
, Zarokanellos, N\BPBI D.%
\BCBL {}\ \BBA {} Fablet, R.%
\end{APACrefauthors}%
\unskip\
\newblock
\APACrefYearMonthDay{2024}{}{}.
\newblock
{\BBOQ}\APACrefatitle {{CLOIN}et: ocean state reconstructions through remote-sensing, in-situ sparse observations and deep learning} {{CLOIN}et: ocean state reconstructions through remote-sensing, in-situ sparse observations and deep learning}.{\BBCQ}
\newblock
\APACjournalVolNumPages{Frontiers in Marine Science}{11}{}{1151868}.
\newblock
\begin{APACrefDOI} \doi{10.3389/fmars.2024.1151868} \end{APACrefDOI}
\PrintBackRefs{\CurrentBib}

\bibitem [\protect \citeauthoryear {%
Dematteis%
, Le~Boyer%
, Pollmann%
, Whalen%
\BCBL {}\ \BBA {} Lvov%
}{%
Dematteis%
\ \protect \BOthers {.}}{%
{\protect \APACyear {2024}}%
}]{%
dematteis2024interacting}
\APACinsertmetastar {%
dematteis2024interacting}%
\begin{APACrefauthors}%
Dematteis, G.%
, Le~Boyer, A.%
, Pollmann, F.%
, Whalen, C\BPBI B.%
\BCBL {}\ \BBA {} Lvov, Y\BPBI V.%
\end{APACrefauthors}%
\unskip\
\newblock
\APACrefYearMonthDay{2024}{}{}.
\newblock
{\BBOQ}\APACrefatitle {Interacting internal waves explain global patterns of interior ocean mixing} {Interacting internal waves explain global patterns of interior ocean mixing}.{\BBCQ}
\newblock
\APACjournalVolNumPages{Nature Communications}{}{}{}.
\newblock
\begin{APACrefDOI} \doi{10.1038/s41467-024-51503-6} \end{APACrefDOI}
\PrintBackRefs{\CurrentBib}

\bibitem [\protect \citeauthoryear {%
Demo%
, Tezzele%
\BCBL {}\ \BBA {} Rozza%
}{%
Demo%
\ \protect \BOthers {.}}{%
{\protect \APACyear {2018}}%
}]{%
demo2018pydmd}
\APACinsertmetastar {%
demo2018pydmd}%
\begin{APACrefauthors}%
Demo, N.%
, Tezzele, M.%
\BCBL {}\ \BBA {} Rozza, G.%
\end{APACrefauthors}%
\unskip\
\newblock
\APACrefYearMonthDay{2018}{}{}.
\newblock
{\BBOQ}\APACrefatitle {{\tt PyDMD}: Python dynamic mode decomposition} {{\tt PyDMD}: Python dynamic mode decomposition}.{\BBCQ}
\newblock
\APACjournalVolNumPages{Journal of Open Source Software}{3}{22}{530}.
\newblock
\begin{APACrefURL} \url{https://github.com/PyDMD/PyDMD?tab=readme-ov-file} \end{APACrefURL}
\newblock
\begin{APACrefDOI} \doi{10.21105/joss.00530} \end{APACrefDOI}
\PrintBackRefs{\CurrentBib}

\bibitem [\protect \citeauthoryear {%
De~Moortel%
, Munday%
\BCBL {}\ \BBA {} Hood%
}{%
De~Moortel%
\ \protect \BOthers {.}}{%
{\protect \APACyear {2004}}%
}]{%
de2004wavelet}
\APACinsertmetastar {%
de2004wavelet}%
\begin{APACrefauthors}%
De~Moortel, I.%
, Munday, S.%
\BCBL {}\ \BBA {} Hood, A\BPBI W.%
\end{APACrefauthors}%
\unskip\
\newblock
\APACrefYearMonthDay{2004}{}{}.
\newblock
{\BBOQ}\APACrefatitle {Wavelet analysis: the effect of varying basic wavelet parameters} {Wavelet analysis: the effect of varying basic wavelet parameters}.{\BBCQ}
\newblock
\APACjournalVolNumPages{Solar Physics}{222}{2}{203--228}.
\newblock
\begin{APACrefDOI} \doi{10.1023/B:SOLA.0000043578.01201.2d} \end{APACrefDOI}
\PrintBackRefs{\CurrentBib}

\bibitem [\protect \citeauthoryear {%
Dewar%
, Parfitt%
\BCBL {}\ \BBA {} Wienders%
}{%
Dewar%
\ \protect \BOthers {.}}{%
{\protect \APACyear {2022}}%
}]{%
dewar2022routine}
\APACinsertmetastar {%
dewar2022routine}%
\begin{APACrefauthors}%
Dewar, W\BPBI K.%
, Parfitt, R.%
\BCBL {}\ \BBA {} Wienders, N.%
\end{APACrefauthors}%
\unskip\
\newblock
\APACrefYearMonthDay{2022}{}{}.
\newblock
{\BBOQ}\APACrefatitle {Routine reversal of the {AMOC} in an ocean model ensemble} {Routine reversal of the {AMOC} in an ocean model ensemble}.{\BBCQ}
\newblock
\APACjournalVolNumPages{Geophysical Research Letters}{49}{24}{e2022GL100117}.
\newblock
\begin{APACrefDOI} \doi{10.1029/2022GL100117} \end{APACrefDOI}
\PrintBackRefs{\CurrentBib}

\bibitem [\protect \citeauthoryear {%
Dibarboure%
\ \protect \BOthers {.}}{%
Dibarboure%
\ \protect \BOthers {.}}{%
{\protect \APACyear {2024}}%
}]{%
dibarboure2024blend}
\APACinsertmetastar {%
dibarboure2024blend}%
\begin{APACrefauthors}%
Dibarboure, G.%
, Anadon, C.%
, Briol, F.%
, Chevrier, R.%
, Delepoulle, A.%
, Faug\`ere, Y.%
\BDBL {}Ubelmann, C.%
\end{APACrefauthors}%
\unskip\
\newblock
\APACrefYearMonthDay{2024}{}{}.
\newblock
{\BBOQ}\APACrefatitle {{Blending 2D topography images from SWOT into the altimeter constellation with the Level-3 multi-mission DUACS system}} {{Blending 2D topography images from SWOT into the altimeter constellation with the Level-3 multi-mission DUACS system}}.{\BBCQ}
\newblock
\APACjournalVolNumPages{EGUsphere}{}{}{}.
\newblock
\begin{APACrefDOI} \doi{10.5194/egusphere-2024-1501} \end{APACrefDOI}
\PrintBackRefs{\CurrentBib}

\bibitem [\protect \citeauthoryear {%
Dylewsky%
, Tao%
\BCBL {}\ \BBA {} Kutz%
}{%
Dylewsky%
\ \protect \BOthers {.}}{%
{\protect \APACyear {2019}}%
}]{%
Dylewsky_2019}
\APACinsertmetastar {%
Dylewsky_2019}%
\begin{APACrefauthors}%
Dylewsky, D.%
, Tao, M.%
\BCBL {}\ \BBA {} Kutz, J\BPBI N.%
\end{APACrefauthors}%
\unskip\
\newblock
\APACrefYearMonthDay{2019}{}{}.
\newblock
{\BBOQ}\APACrefatitle {Dynamic mode decomposition for multiscale nonlinear physics} {Dynamic mode decomposition for multiscale nonlinear physics}.{\BBCQ}
\newblock
\APACjournalVolNumPages{Physical Review E}{99}{6}{}.
\newblock
\begin{APACrefDOI} \doi{10.1103/physreve.99.063311} \end{APACrefDOI}
\PrintBackRefs{\CurrentBib}

\bibitem [\protect \citeauthoryear {%
Early%
, Avila%
, Fabre-Lima%
\BCBL {}\ \BBA {} Sundermeyer%
}{%
Early%
, Avila%
\BCBL {}\ \protect \BOthers {.}}{%
{\protect \APACyear {2024}}%
}]{%
early2024wvm}
\APACinsertmetastar {%
early2024wvm}%
\begin{APACrefauthors}%
Early, J\BPBI J.%
, Avila, B.%
, Fabre-Lima, L.%
\BCBL {}\ \BBA {} Sundermeyer, M\BPBI A.%
\end{APACrefauthors}%
\unskip\
\newblock
\APACrefYearMonthDay{2024}{{\APACmonth{07}}}{}.
\newblock
\APACrefbtitle {{{\tt Energy-Pathways-Group/GLOceanKit}: Pre-release of the non-hydrostatic wave-vortex model [Software]}.} {{{\tt Energy-Pathways-Group/GLOceanKit}: Pre-release of the non-hydrostatic wave-vortex model [Software]}.}
\newblock
\APACaddressPublisher{}{Zenodo}.
\newblock
\begin{APACrefURL} \url{https://doi.org/10.5281/zenodo.12839943} \end{APACrefURL}
\newblock
\begin{APACrefDOI} \doi{10.5281/zenodo.12839943} \end{APACrefDOI}
\PrintBackRefs{\CurrentBib}

\bibitem [\protect \citeauthoryear {%
Early%
, Hern{\'a}ndez-Due{\~n}as%
, Smith%
, Lelong%
\BCBL {}\ \protect \BOthers {.}}{%
Early%
, Hern{\'a}ndez-Due{\~n}as%
\BCBL {}\ \protect \BOthers {.}}{%
{\protect \APACyear {2024}}%
}]{%
early2024available}
\APACinsertmetastar {%
early2024available}%
\begin{APACrefauthors}%
Early, J\BPBI J.%
, Hern{\'a}ndez-Due{\~n}as, G.%
, Smith, L\BPBI M.%
, Lelong, M\BPBI P.%
\BCBL {}\ \BOthersPeriod {.}\end{APACrefauthors}%
\unskip\
\newblock
\APACrefYearMonthDay{2024}{}{}.
\newblock
{\BBOQ}\APACrefatitle {Available potential vorticity and the wave-vortex decomposition for arbitrary stratification} {Available potential vorticity and the wave-vortex decomposition for arbitrary stratification}.{\BBCQ}
\newblock
\APACjournalVolNumPages{arXiv preprint arXiv:2403.20269}{}{}{}.
\newblock
\begin{APACrefDOI} \doi{10.48550/arXiv.2403.20269} \end{APACrefDOI}
\PrintBackRefs{\CurrentBib}

\bibitem [\protect \citeauthoryear {%
Early%
, Lelong%
\BCBL {}\ \BBA {} Sundermeyer%
}{%
Early%
\ \protect \BOthers {.}}{%
{\protect \APACyear {2021}}%
}]{%
early2021generalized}
\APACinsertmetastar {%
early2021generalized}%
\begin{APACrefauthors}%
Early, J\BPBI J.%
, Lelong, M\BPBI P.%
\BCBL {}\ \BBA {} Sundermeyer, M\BPBI A.%
\end{APACrefauthors}%
\unskip\
\newblock
\APACrefYearMonthDay{2021}{}{}.
\newblock
{\BBOQ}\APACrefatitle {A generalized wave-vortex decomposition for rotating {B}oussinesq flows with arbitrary stratification} {A generalized wave-vortex decomposition for rotating {B}oussinesq flows with arbitrary stratification}.{\BBCQ}
\newblock
\APACjournalVolNumPages{Journal of Fluid Mechanics}{912}{}{A32}.
\newblock
\begin{APACrefDOI} \doi{10.1017/jfm.2020.995} \end{APACrefDOI}
\PrintBackRefs{\CurrentBib}

\bibitem [\protect \citeauthoryear {%
Fablet%
, Chapron%
, Le~Sommer%
\BCBL {}\ \BBA {} S{\'e}vellec%
}{%
Fablet%
\ \protect \BOthers {.}}{%
{\protect \APACyear {2024}}%
}]{%
fablet2024inversion}
\APACinsertmetastar {%
fablet2024inversion}%
\begin{APACrefauthors}%
Fablet, R.%
, Chapron, B.%
, Le~Sommer, J.%
\BCBL {}\ \BBA {} S{\'e}vellec, F.%
\end{APACrefauthors}%
\unskip\
\newblock
\APACrefYearMonthDay{2024}{}{}.
\newblock
{\BBOQ}\APACrefatitle {{Inversion of sea surface currents from satellite-derived SST-SSH synergies with 4DVarNets}} {{Inversion of sea surface currents from satellite-derived SST-SSH synergies with 4DVarNets}}.{\BBCQ}
\newblock
\APACjournalVolNumPages{Journal of Advances in Modeling Earth Systems}{}{}{}.
\newblock
\begin{APACrefDOI} \doi{10.1029/2023MS003609} \end{APACrefDOI}
\PrintBackRefs{\CurrentBib}

\bibitem [\protect \citeauthoryear {%
Febvre%
, Le~Sommer%
, Ubelmann%
\BCBL {}\ \BBA {} Fablet%
}{%
Febvre%
\ \protect \BOthers {.}}{%
{\protect \APACyear {2024}}%
}]{%
febvre2024training}
\APACinsertmetastar {%
febvre2024training}%
\begin{APACrefauthors}%
Febvre, Q.%
, Le~Sommer, J.%
, Ubelmann, C.%
\BCBL {}\ \BBA {} Fablet, R.%
\end{APACrefauthors}%
\unskip\
\newblock
\APACrefYearMonthDay{2024}{}{}.
\newblock
{\BBOQ}\APACrefatitle {Training neural mapping schemes for satellite altimetry with simulation data} {Training neural mapping schemes for satellite altimetry with simulation data}.{\BBCQ}
\newblock
\APACjournalVolNumPages{Journal of Advances in Modeling Earth Systems}{16}{7}{}.
\newblock
\begin{APACrefDOI} \doi{10.1029/2023MS003959} \end{APACrefDOI}
\PrintBackRefs{\CurrentBib}

\bibitem [\protect \citeauthoryear {%
Gao%
\ \protect \BOthers {.}}{%
Gao%
\ \protect \BOthers {.}}{%
{\protect \APACyear {2024}}%
}]{%
gao2024deep}
\APACinsertmetastar {%
gao2024deep}%
\begin{APACrefauthors}%
Gao, Z.%
, Chapron, B.%
, Ma, C.%
, Fablet, R.%
, Febvre, Q.%
, Zhao, W.%
\BCBL {}\ \BBA {} Chen, G.%
\end{APACrefauthors}%
\unskip\
\newblock
\APACrefYearMonthDay{2024}{}{}.
\newblock
{\BBOQ}\APACrefatitle {A deep learning approach to extract balanced motions from sea surface height snapshot} {A deep learning approach to extract balanced motions from sea surface height snapshot}.{\BBCQ}
\newblock
\APACjournalVolNumPages{Geophysical Research Letters}{51}{7}{e2023GL106623}.
\newblock
\begin{APACrefDOI} \doi{10.1029/2023GL106623} \end{APACrefDOI}
\PrintBackRefs{\CurrentBib}

\bibitem [\protect \citeauthoryear {%
Garrett%
\ \BBA {} Munk%
}{%
Garrett%
\ \BBA {} Munk%
}{%
{\protect \APACyear {1975}}%
}]{%
garrett1975space}
\APACinsertmetastar {%
garrett1975space}%
\begin{APACrefauthors}%
Garrett, C.%
\BCBT {}\ \BBA {} Munk, W.%
\end{APACrefauthors}%
\unskip\
\newblock
\APACrefYearMonthDay{1975}{}{}.
\newblock
{\BBOQ}\APACrefatitle {Space-time scales of internal waves: A progress report} {Space-time scales of internal waves: A progress report}.{\BBCQ}
\newblock
\APACjournalVolNumPages{Journal of Geophysical Research}{80}{3}{291--297}.
\newblock
\begin{APACrefDOI} \doi{10.1029/JC080i003p00291} \end{APACrefDOI}
\PrintBackRefs{\CurrentBib}

\bibitem [\protect \citeauthoryear {%
Grooms%
\ \protect \BOthers {.}}{%
Grooms%
\ \protect \BOthers {.}}{%
{\protect \APACyear {2021}}%
}]{%
grooms2021diffusion}
\APACinsertmetastar {%
grooms2021diffusion}%
\begin{APACrefauthors}%
Grooms, I.%
, Loose, N.%
, Abernathey, R\BPBI P.%
, Steinberg, J\BPBI M.%
, Bachman, S\BPBI D.%
, Marques, G.%
\BDBL {}Yankovsky, E.%
\end{APACrefauthors}%
\unskip\
\newblock
\APACrefYearMonthDay{2021}{}{}.
\newblock
{\BBOQ}\APACrefatitle {Diffusion-based smoothers for spatial filtering of gridded geophysical data} {Diffusion-based smoothers for spatial filtering of gridded geophysical data}.{\BBCQ}
\newblock
\APACjournalVolNumPages{Journal of Advances in Modeling Earth Systems}{13}{9}{e2021MS002552}.
\newblock
\begin{APACrefDOI} \doi{10.1029/2021MS002552} \end{APACrefDOI}
\PrintBackRefs{\CurrentBib}

\bibitem [\protect \citeauthoryear {%
Hiron%
, Nolan%
\BCBL {}\ \BBA {} Shay%
}{%
Hiron%
\ \protect \BOthers {.}}{%
{\protect \APACyear {2021}}%
}]{%
Hironetal_2021}
\APACinsertmetastar {%
Hironetal_2021}%
\begin{APACrefauthors}%
Hiron, L.%
, Nolan, D\BPBI S.%
\BCBL {}\ \BBA {} Shay, L\BPBI K.%
\end{APACrefauthors}%
\unskip\
\newblock
\APACrefYearMonthDay{2021}{}{}.
\newblock
{\BBOQ}\APACrefatitle {{Study of Ageostrophy during Strong, Nonlinear Eddy-Front Interaction in the Gulf of Mexico}} {{Study of Ageostrophy during Strong, Nonlinear Eddy-Front Interaction in the Gulf of Mexico}}.{\BBCQ}
\newblock
\APACjournalVolNumPages{Journal of Physical Oceanography}{51}{3}{745 - 755}.
\newblock
\begin{APACrefDOI} \doi{10.1175/JPO-D-20-0182.1} \end{APACrefDOI}
\PrintBackRefs{\CurrentBib}

\bibitem [\protect \citeauthoryear {%
Ichinaga%
\ \protect \BOthers {.}}{%
Ichinaga%
\ \protect \BOthers {.}}{%
{\protect \APACyear {2024}}%
}]{%
ichinaga2024pydmd}
\APACinsertmetastar {%
ichinaga2024pydmd}%
\begin{APACrefauthors}%
Ichinaga, S\BPBI M.%
, Andreuzzi, F.%
, Demo, N.%
, Tezzele, M.%
, Lapo, K.%
, Rozza, G.%
\BDBL {}Kutz, J\BPBI N.%
\end{APACrefauthors}%
\unskip\
\newblock
\APACrefYearMonthDay{2024}{}{}.
\newblock
{\BBOQ}\APACrefatitle {{{\tt PyDMD}: A Python package for robust dynamic mode decomposition}} {{{\tt PyDMD}: A Python package for robust dynamic mode decomposition}}.{\BBCQ}
\newblock
\APACjournalVolNumPages{arXiv preprint arXiv:2402.07463}{}{}{}.
\newblock
\begin{APACrefURL} \url{https://pydmd.github.io/PyDMD/} \end{APACrefURL}
\newblock
\begin{APACrefDOI} \doi{10.48550/arXiv.2402.07463} \end{APACrefDOI}
\PrintBackRefs{\CurrentBib}

\bibitem [\protect \citeauthoryear {%
Jones%
, Xiao%
, Abernathey%
\BCBL {}\ \BBA {} Smith%
}{%
Jones%
\ \protect \BOthers {.}}{%
{\protect \APACyear {2023}}%
}]{%
jones2023using}
\APACinsertmetastar {%
jones2023using}%
\begin{APACrefauthors}%
Jones, C\BPBI S.%
, Xiao, Q.%
, Abernathey, R\BPBI P.%
\BCBL {}\ \BBA {} Smith, K\BPBI S.%
\end{APACrefauthors}%
\unskip\
\newblock
\APACrefYearMonthDay{2023}{}{}.
\newblock
{\BBOQ}\APACrefatitle {{Using Lagrangian filtering to remove waves from the ocean surface velocity field}} {{Using Lagrangian filtering to remove waves from the ocean surface velocity field}}.{\BBCQ}
\newblock
\APACjournalVolNumPages{Journal of Advances in Modeling Earth Systems}{15}{4}{e2022MS003220}.
\newblock
\begin{APACrefDOI} \doi{10.1029/2022MS003220} \end{APACrefDOI}
\PrintBackRefs{\CurrentBib}

\bibitem [\protect \citeauthoryear {%
Kutz%
, Brunton%
, Brunton%
\BCBL {}\ \BBA {} Proctor%
}{%
Kutz%
\ \protect \BOthers {.}}{%
{\protect \APACyear {2016}}%
}]{%
kutz2016dynamic}
\APACinsertmetastar {%
kutz2016dynamic}%
\begin{APACrefauthors}%
Kutz, J\BPBI N.%
, Brunton, S\BPBI L.%
, Brunton, B\BPBI W.%
\BCBL {}\ \BBA {} Proctor, J\BPBI L.%
\end{APACrefauthors}%
\unskip\
\newblock
\APACrefYear{2016}.
\newblock
\APACrefbtitle {Dynamic mode decomposition: Data-driven modeling of complex systems} {Dynamic mode decomposition: Data-driven modeling of complex systems}.
\newblock
\APACaddressPublisher{}{SIAM}.
\newblock
\begin{APACrefDOI} \doi{10.1137/1.9781611974508} \end{APACrefDOI}
\PrintBackRefs{\CurrentBib}

\bibitem [\protect \citeauthoryear {%
Kutz%
, Reza%
, Faraji%
\BCBL {}\ \BBA {} Knoll%
}{%
Kutz%
\ \protect \BOthers {.}}{%
{\protect \APACyear {2024}}%
}]{%
kutz2024shallow}
\APACinsertmetastar {%
kutz2024shallow}%
\begin{APACrefauthors}%
Kutz, J\BPBI N.%
, Reza, M.%
, Faraji, F.%
\BCBL {}\ \BBA {} Knoll, A.%
\end{APACrefauthors}%
\unskip\
\newblock
\APACrefYearMonthDay{2024}{}{}.
\newblock
{\BBOQ}\APACrefatitle {Shallow Recurrent Decoder for Reduced Order Modeling of Plasma Dynamics} {Shallow recurrent decoder for reduced order modeling of plasma dynamics}.{\BBCQ}
\newblock
\APACjournalVolNumPages{arXiv preprint arXiv:2405.11955}{}{}{}.
\newblock
\begin{APACrefDOI} \doi{10.48550/arXiv.2405.11955} \end{APACrefDOI}
\PrintBackRefs{\CurrentBib}

\bibitem [\protect \citeauthoryear {%
Lapo%
, Ichinaga%
\BCBL {}\ \BBA {} Kutz%
}{%
Lapo%
\ \protect \BOthers {.}}{%
{\protect \APACyear {2024}}%
}]{%
lapo2024multi}
\APACinsertmetastar {%
lapo2024multi}%
\begin{APACrefauthors}%
Lapo, K.%
, Ichinaga, S\BPBI M.%
\BCBL {}\ \BBA {} Kutz, J\BPBI N.%
\end{APACrefauthors}%
\unskip\
\newblock
\APACrefYearMonthDay{2024}{}{}.
\newblock
{\BBOQ}\APACrefatitle {Unsupervised multi-scale diagnostics} {Unsupervised multi-scale diagnostics}.{\BBCQ}
\newblock
\APACjournalVolNumPages{arXiv preprint arXiv:2408.02396}{}{}{}.
\newblock
\begin{APACrefDOI} \doi{10.48550/arXiv:2408.02396} \end{APACrefDOI}
\PrintBackRefs{\CurrentBib}

\bibitem [\protect \citeauthoryear {%
Le~Guillou%
\ \protect \BOthers {.}}{%
Le~Guillou%
\ \protect \BOthers {.}}{%
{\protect \APACyear {2023}}%
}]{%
le2023regional}
\APACinsertmetastar {%
le2023regional}%
\begin{APACrefauthors}%
Le~Guillou, F.%
, Gaultier, L.%
, Ballarotta, M.%
, Metref, S.%
, Ubelmann, C.%
, Cosme, E.%
\BCBL {}\ \BBA {} Rio, M\BHBI H.%
\end{APACrefauthors}%
\unskip\
\newblock
\APACrefYearMonthDay{2023}{}{}.
\newblock
{\BBOQ}\APACrefatitle {Regional mapping of energetic short mesoscale ocean dynamics from altimetry: performances from real observations} {Regional mapping of energetic short mesoscale ocean dynamics from altimetry: performances from real observations}.{\BBCQ}
\newblock
\APACjournalVolNumPages{Ocean Science}{19}{5}{1517--1527}.
\newblock
\begin{APACrefDOI} \doi{10.5194/os-19-1517-2023} \end{APACrefDOI}
\PrintBackRefs{\CurrentBib}

\bibitem [\protect \citeauthoryear {%
L.~Li%
, Deremble%
, Lahaye%
\BCBL {}\ \BBA {} M{\'e}min%
}{%
L.~Li%
\ \protect \BOthers {.}}{%
{\protect \APACyear {2023}}%
}]{%
li2023stochastic}
\APACinsertmetastar {%
li2023stochastic}%
\begin{APACrefauthors}%
Li, L.%
, Deremble, B.%
, Lahaye, N.%
\BCBL {}\ \BBA {} M{\'e}min, E.%
\end{APACrefauthors}%
\unskip\
\newblock
\APACrefYearMonthDay{2023}{}{}.
\newblock
{\BBOQ}\APACrefatitle {Stochastic data-driven parameterization of unresolved eddy effects in a baroclinic quasi-geostrophic model} {Stochastic data-driven parameterization of unresolved eddy effects in a baroclinic quasi-geostrophic model}.{\BBCQ}
\newblock
\APACjournalVolNumPages{Journal of Advances in Modeling Earth Systems}{15}{2}{e2022MS003297}.
\newblock
\begin{APACrefDOI} \doi{10.1029/2022MS003297} \end{APACrefDOI}
\PrintBackRefs{\CurrentBib}

\bibitem [\protect \citeauthoryear {%
Y.~Li%
, Huang%
, Ma%
, Feng%
\BCBL {}\ \BBA {} Liang%
}{%
Y.~Li%
\ \protect \BOthers {.}}{%
{\protect \APACyear {2024}}%
}]{%
li2024snow}
\APACinsertmetastar {%
li2024snow}%
\begin{APACrefauthors}%
Li, Y.%
, Huang, X.%
, Ma, Y.%
, Feng, Q.%
\BCBL {}\ \BBA {} Liang, T.%
\end{APACrefauthors}%
\unskip\
\newblock
\APACrefYearMonthDay{2024}{}{}.
\newblock
{\BBOQ}\APACrefatitle {Snow Drought Patterns and Their Spatiotemporal Heterogeneity in {C}hina} {Snow drought patterns and their spatiotemporal heterogeneity in {C}hina}.{\BBCQ}
\newblock
\APACjournalVolNumPages{IEEE Journal of Selected Topics in Applied Earth Observations and Remote Sensing}{17}{}{2029--2036}.
\newblock
\begin{APACrefDOI} \doi{10.1109/JSTARS.2023.3344763} \end{APACrefDOI}
\PrintBackRefs{\CurrentBib}

\bibitem [\protect \citeauthoryear {%
Loose%
\ \protect \BOthers {.}}{%
Loose%
\ \protect \BOthers {.}}{%
{\protect \APACyear {2022}}%
}]{%
loose2022gcm}
\APACinsertmetastar {%
loose2022gcm}%
\begin{APACrefauthors}%
Loose, N.%
, Abernathey, R.%
, Grooms, I.%
, Busecke, J\BPBI J\BPBI M.%
, Guillaumin, A.%
, Yankovsky, E.%
\BDBL {}others%
\end{APACrefauthors}%
\unskip\
\newblock
\APACrefYearMonthDay{2022}{}{}.
\newblock
{\BBOQ}\APACrefatitle {{\tt{GCM}-filters}: A Python package for diffusion-based spatial filtering of gridded data} {{\tt{GCM}-filters}: A python package for diffusion-based spatial filtering of gridded data}.{\BBCQ}
\newblock
\APACjournalVolNumPages{Journal of Open Source Software}{7}{70}{}.
\newblock
\begin{APACrefDOI} \doi{10.21105/joss.03947} \end{APACrefDOI}
\PrintBackRefs{\CurrentBib}

\bibitem [\protect \citeauthoryear {%
Lyu%
, Wang%
, Pedersen%
, Jones%
\BCBL {}\ \BBA {} Balwada%
}{%
Lyu%
\ \protect \BOthers {.}}{%
{\protect \APACyear {2024}}%
}]{%
lyu2024multi}
\APACinsertmetastar {%
lyu2024multi}%
\begin{APACrefauthors}%
Lyu, J.%
, Wang, Y.%
, Pedersen, C.%
, Jones, C\BPBI S.%
\BCBL {}\ \BBA {} Balwada, D.%
\end{APACrefauthors}%
\unskip\
\newblock
\APACrefYearMonthDay{2024}{}{}.
\newblock
{\BBOQ}\APACrefatitle {Multi-scale decomposition of sea surface height snapshots using machine learning} {Multi-scale decomposition of sea surface height snapshots using machine learning}.{\BBCQ}
\newblock
\APACjournalVolNumPages{arXiv preprint arXiv:2409.17354}{}{}{}.
\newblock
\begin{APACrefDOI} \doi{10.48550/arXiv.2409.17354} \end{APACrefDOI}
\PrintBackRefs{\CurrentBib}

\bibitem [\protect \citeauthoryear {%
Maingonnat%
, Tissot%
\BCBL {}\ \BBA {} Lahaye%
}{%
Maingonnat%
\ \protect \BOthers {.}}{%
{\protect \APACyear {2024}}%
}]{%
Maingonnat_2024}
\APACinsertmetastar {%
Maingonnat_2024}%
\begin{APACrefauthors}%
Maingonnat, I.%
, Tissot, G.%
\BCBL {}\ \BBA {} Lahaye, N.%
\end{APACrefauthors}%
\unskip\
\newblock
\APACrefYearMonthDay{2024}{}{}.
\newblock
{\BBOQ}\APACrefatitle {Coupled estimation of incoherent inertia gravity wave field and turbulent balanced motions via modal decomposition} {Coupled estimation of incoherent inertia gravity wave field and turbulent balanced motions via modal decomposition}.{\BBCQ}
\newblock
\APACjournalVolNumPages{EGUsphere}{}{}{1--31}.
\newblock
\begin{APACrefDOI} \doi{10.5194/egusphere-2024-1483} \end{APACrefDOI}
\PrintBackRefs{\CurrentBib}

\bibitem [\protect \citeauthoryear {%
Martin%
, Manucharyan%
\BCBL {}\ \BBA {} Klein%
}{%
Martin%
\ \protect \BOthers {.}}{%
{\protect \APACyear {2024}}%
}]{%
martin2024deep}
\APACinsertmetastar {%
martin2024deep}%
\begin{APACrefauthors}%
Martin, S\BPBI A.%
, Manucharyan, G.%
\BCBL {}\ \BBA {} Klein, P.%
\end{APACrefauthors}%
\unskip\
\newblock
\APACrefYearMonthDay{2024}{}{}.
\newblock
{\BBOQ}\APACrefatitle {Deep Learning Improves Global Satellite Observations of Ocean Eddy Dynamics} {Deep learning improves global satellite observations of ocean eddy dynamics}.{\BBCQ}
\newblock
\APACjournalVolNumPages{Geophysical Research Letters}{51}{17}{}.
\newblock
\begin{APACrefDOI} \doi{10.1029/2024GL110059} \end{APACrefDOI}
\PrintBackRefs{\CurrentBib}

\bibitem [\protect \citeauthoryear {%
McWilliams%
}{%
McWilliams%
}{%
{\protect \APACyear {2019}}%
}]{%
mcwilliams2019survey}
\APACinsertmetastar {%
mcwilliams2019survey}%
\begin{APACrefauthors}%
McWilliams, J\BPBI C.%
\end{APACrefauthors}%
\unskip\
\newblock
\APACrefYearMonthDay{2019}{}{}.
\newblock
{\BBOQ}\APACrefatitle {A survey of submesoscale currents} {A survey of submesoscale currents}.{\BBCQ}
\newblock
\APACjournalVolNumPages{Geoscience Letters}{6}{1}{1--15}.
\newblock
\begin{APACrefDOI} \doi{10.1186/s40562-019-0133-3} \end{APACrefDOI}
\PrintBackRefs{\CurrentBib}

\bibitem [\protect \citeauthoryear {%
McWilliams%
}{%
McWilliams%
}{%
{\protect \APACyear {2021}}%
}]{%
mcwilliams2021oceanic}
\APACinsertmetastar {%
mcwilliams2021oceanic}%
\begin{APACrefauthors}%
McWilliams, J\BPBI C.%
\end{APACrefauthors}%
\unskip\
\newblock
\APACrefYearMonthDay{2021}{}{}.
\newblock
{\BBOQ}\APACrefatitle {Oceanic frontogenesis} {Oceanic frontogenesis}.{\BBCQ}
\newblock
\APACjournalVolNumPages{Annual Review of Marine Science}{13}{1}{227--253}.
\newblock
\begin{APACrefDOI} \doi{10.1146/annurev-marine-032320-120725} \end{APACrefDOI}
\PrintBackRefs{\CurrentBib}

\bibitem [\protect \citeauthoryear {%
Minz%
, Baker%
, Kafiabad%
\BCBL {}\ \BBA {} Vanneste%
}{%
Minz%
\ \protect \BOthers {.}}{%
{\protect \APACyear {2024}}%
}]{%
minz2024exponential}
\APACinsertmetastar {%
minz2024exponential}%
\begin{APACrefauthors}%
Minz, A.%
, Baker, L\BPBI E.%
, Kafiabad, H\BPBI A.%
\BCBL {}\ \BBA {} Vanneste, J.%
\end{APACrefauthors}%
\unskip\
\newblock
\APACrefYearMonthDay{2024}{}{}.
\newblock
{\BBOQ}\APACrefatitle {The exponential Lagrangian mean} {The exponential lagrangian mean}.{\BBCQ}
\newblock
\APACjournalVolNumPages{arXiv preprint arXiv:2406.18243}{}{}{}.
\PrintBackRefs{\CurrentBib}

\bibitem [\protect \citeauthoryear {%
Mishonov%
, Seidov%
\BCBL {}\ \BBA {} Reagan%
}{%
Mishonov%
\ \protect \BOthers {.}}{%
{\protect \APACyear {2024}}%
}]{%
mishonov2024revisiting}
\APACinsertmetastar {%
mishonov2024revisiting}%
\begin{APACrefauthors}%
Mishonov, A.%
, Seidov, D.%
\BCBL {}\ \BBA {} Reagan, J.%
\end{APACrefauthors}%
\unskip\
\newblock
\APACrefYearMonthDay{2024}{}{}.
\newblock
{\BBOQ}\APACrefatitle {Revisiting the multidecadal variability of {North Atlantic Ocean} circulation and climate} {Revisiting the multidecadal variability of {North Atlantic Ocean} circulation and climate}.{\BBCQ}
\newblock
\APACjournalVolNumPages{Frontiers in Marine Science}{11}{}{1345426}.
\newblock
\begin{APACrefDOI} \doi{10.3389/fmars.2024.1345426} \end{APACrefDOI}
\PrintBackRefs{\CurrentBib}

\bibitem [\protect \citeauthoryear {%
Miyamoto%
\ \BBA {} Xie%
}{%
Miyamoto%
\ \BBA {} Xie%
}{%
{\protect \APACyear {2024}}%
}]{%
miyamoto2024low}
\APACinsertmetastar {%
miyamoto2024low}%
\begin{APACrefauthors}%
Miyamoto, A.%
\BCBT {}\ \BBA {} Xie, S\BHBI P.%
\end{APACrefauthors}%
\unskip\
\newblock
\APACrefYearMonthDay{2024}{}{}.
\newblock
{\BBOQ}\APACrefatitle {{Low cloud-SST variability over the summertime subtropical Northeast Pacific: Role of extratropical atmospheric modes}} {{Low cloud-SST variability over the summertime subtropical Northeast Pacific: Role of extratropical atmospheric modes}}.{\BBCQ}
\newblock
\APACjournalVolNumPages{Journal of Climate}{}{}{}.
\newblock
\begin{APACrefDOI} \doi{10.1175/JCLI-D-24-0015.1} \end{APACrefDOI}
\PrintBackRefs{\CurrentBib}

\bibitem [\protect \citeauthoryear {%
Mojgani%
, Chattopadhyay%
\BCBL {}\ \BBA {} Hassanzadeh%
}{%
Mojgani%
\ \protect \BOthers {.}}{%
{\protect \APACyear {2024}}%
}]{%
mojgani2024interpretable}
\APACinsertmetastar {%
mojgani2024interpretable}%
\begin{APACrefauthors}%
Mojgani, R.%
, Chattopadhyay, A.%
\BCBL {}\ \BBA {} Hassanzadeh, P.%
\end{APACrefauthors}%
\unskip\
\newblock
\APACrefYearMonthDay{2024}{}{}.
\newblock
{\BBOQ}\APACrefatitle {Interpretable structural model error discovery from sparse assimilation increments using spectral bias-reduced neural networks: A quasi-geostrophic turbulence test case} {Interpretable structural model error discovery from sparse assimilation increments using spectral bias-reduced neural networks: A quasi-geostrophic turbulence test case}.{\BBCQ}
\newblock
\APACjournalVolNumPages{Journal of Advances in Modeling Earth Systems}{16}{3}{e2023MS004033}.
\newblock
\begin{APACrefDOI} \doi{10.1029/2023MS004033} \end{APACrefDOI}
\PrintBackRefs{\CurrentBib}

\bibitem [\protect \citeauthoryear {%
Newman%
\ \protect \BOthers {.}}{%
Newman%
\ \protect \BOthers {.}}{%
{\protect \APACyear {2016}}%
}]{%
newman2016pacific}
\APACinsertmetastar {%
newman2016pacific}%
\begin{APACrefauthors}%
Newman, M.%
, Alexander, M\BPBI A.%
, Ault, T\BPBI R.%
, Cobb, K\BPBI M.%
, Deser, C.%
, Di~Lorenzo, E.%
\BDBL {}others%
\end{APACrefauthors}%
\unskip\
\newblock
\APACrefYearMonthDay{2016}{}{}.
\newblock
{\BBOQ}\APACrefatitle {{The Pacific Decadal Oscillation, revisited}} {{The Pacific Decadal Oscillation, revisited}}.{\BBCQ}
\newblock
\APACjournalVolNumPages{Journal of Climate}{29}{12}{4399--4427}.
\newblock
\begin{APACrefDOI} \doi{10.1175/JCLI-D-15-0508.1} \end{APACrefDOI}
\PrintBackRefs{\CurrentBib}

\bibitem [\protect \citeauthoryear {%
Pedlosky%
}{%
Pedlosky%
}{%
{\protect \APACyear {1984}}%
}]{%
pedlosky1984equations}
\APACinsertmetastar {%
pedlosky1984equations}%
\begin{APACrefauthors}%
Pedlosky, J.%
\end{APACrefauthors}%
\unskip\
\newblock
\APACrefYearMonthDay{1984}{}{}.
\newblock
{\BBOQ}\APACrefatitle {The equations for geostrophic motion in the ocean} {The equations for geostrophic motion in the ocean}.{\BBCQ}
\newblock
\APACjournalVolNumPages{Journal of Physical Oceanography}{14}{2}{448--455}.
\newblock
\begin{APACrefDOI} \doi{10.1175/1520-0485(1984)014<0448:TEFGMI>2.0.CO;2} \end{APACrefDOI}
\PrintBackRefs{\CurrentBib}

\bibitem [\protect \citeauthoryear {%
Pedregosa%
\ \protect \BOthers {.}}{%
Pedregosa%
\ \protect \BOthers {.}}{%
{\protect \APACyear {2011}}%
}]{%
scikit-learn}
\APACinsertmetastar {%
scikit-learn}%
\begin{APACrefauthors}%
Pedregosa, F.%
, Varoquaux, G.%
, Gramfort, A.%
, Michel, V.%
, Thirion, B.%
, Grisel, O.%
\BDBL {}Duchesnay, E.%
\end{APACrefauthors}%
\unskip\
\newblock
\APACrefYearMonthDay{2011}{}{}.
\newblock
{\BBOQ}\APACrefatitle {Scikit-learn: Machine Learning in {P}ython} {Scikit-learn: Machine learning in {P}ython}.{\BBCQ}
\newblock
\APACjournalVolNumPages{Journal of Machine Learning Research}{12}{}{2825--2830}.
\newblock
\begin{APACrefURL} \url{https://scikit-learn.org/1.5/modules/clustering.html#k-means} \end{APACrefURL}
\PrintBackRefs{\CurrentBib}

\bibitem [\protect \citeauthoryear {%
Proctor%
\ \BBA {} Eckhoff%
}{%
Proctor%
\ \BBA {} Eckhoff%
}{%
{\protect \APACyear {2015}}%
}]{%
proctor2015discovering}
\APACinsertmetastar {%
proctor2015discovering}%
\begin{APACrefauthors}%
Proctor, J\BPBI L.%
\BCBT {}\ \BBA {} Eckhoff, P\BPBI A.%
\end{APACrefauthors}%
\unskip\
\newblock
\APACrefYearMonthDay{2015}{}{}.
\newblock
{\BBOQ}\APACrefatitle {Discovering dynamic patterns from infectious disease data using dynamic mode decomposition} {Discovering dynamic patterns from infectious disease data using dynamic mode decomposition}.{\BBCQ}
\newblock
\APACjournalVolNumPages{International health}{7}{2}{139--145}.
\newblock
\begin{APACrefDOI} \doi{10.1093/inthealth/ihv009} \end{APACrefDOI}
\PrintBackRefs{\CurrentBib}

\bibitem [\protect \citeauthoryear {%
Richman%
, Arbic%
, Shriver%
, Metzger%
\BCBL {}\ \BBA {} Wallcraft%
}{%
Richman%
\ \protect \BOthers {.}}{%
{\protect \APACyear {2012}}%
}]{%
richman2012inferring}
\APACinsertmetastar {%
richman2012inferring}%
\begin{APACrefauthors}%
Richman, J\BPBI G.%
, Arbic, B\BPBI K.%
, Shriver, J\BPBI F.%
, Metzger, E\BPBI J.%
\BCBL {}\ \BBA {} Wallcraft, A\BPBI J.%
\end{APACrefauthors}%
\unskip\
\newblock
\APACrefYearMonthDay{2012}{}{}.
\newblock
{\BBOQ}\APACrefatitle {Inferring dynamics from the wavenumber spectra of an eddying global ocean model with embedded tides} {Inferring dynamics from the wavenumber spectra of an eddying global ocean model with embedded tides}.{\BBCQ}
\newblock
\APACjournalVolNumPages{Journal of Geophysical Research: Oceans}{117}{C12}{}.
\newblock
\begin{APACrefDOI} \doi{10.1029/2012JC008364} \end{APACrefDOI}
\PrintBackRefs{\CurrentBib}

\bibitem [\protect \citeauthoryear {%
Savage%
\ \protect \BOthers {.}}{%
Savage%
\ \protect \BOthers {.}}{%
{\protect \APACyear {2017}}%
}]{%
savage2017frequency}
\APACinsertmetastar {%
savage2017frequency}%
\begin{APACrefauthors}%
Savage, A\BPBI C.%
, Arbic, B\BPBI K.%
, Richman, J\BPBI G.%
, Shriver, J\BPBI F.%
, Alford, M\BPBI H.%
, Buijsman, M\BPBI C.%
\BDBL {}others%
\end{APACrefauthors}%
\unskip\
\newblock
\APACrefYearMonthDay{2017}{}{}.
\newblock
{\BBOQ}\APACrefatitle {Frequency content of sea surface height variability from internal gravity waves to mesoscale eddies} {Frequency content of sea surface height variability from internal gravity waves to mesoscale eddies}.{\BBCQ}
\newblock
\APACjournalVolNumPages{Journal of Geophysical Research: Oceans}{122}{3}{2519--2538}.
\newblock
\begin{APACrefDOI} \doi{10.1002/2016JC012331} \end{APACrefDOI}
\PrintBackRefs{\CurrentBib}

\bibitem [\protect \citeauthoryear {%
Schmid%
}{%
Schmid%
}{%
{\protect \APACyear {2022}}%
}]{%
schmid2022dynamic}
\APACinsertmetastar {%
schmid2022dynamic}%
\begin{APACrefauthors}%
Schmid, P\BPBI J.%
\end{APACrefauthors}%
\unskip\
\newblock
\APACrefYearMonthDay{2022}{}{}.
\newblock
{\BBOQ}\APACrefatitle {Dynamic mode decomposition and its variants} {Dynamic mode decomposition and its variants}.{\BBCQ}
\newblock
\APACjournalVolNumPages{Annual Review of Fluid Mechanics}{54}{}{225--254}.
\newblock
\begin{APACrefDOI} \doi{10.1146/annurev-fluid-030121-015835} \end{APACrefDOI}
\PrintBackRefs{\CurrentBib}

\bibitem [\protect \citeauthoryear {%
Shakespeare%
\ \protect \BOthers {.}}{%
Shakespeare%
\ \protect \BOthers {.}}{%
{\protect \APACyear {2021}}%
}]{%
shakespeare2021new}
\APACinsertmetastar {%
shakespeare2021new}%
\begin{APACrefauthors}%
Shakespeare, C\BPBI J.%
, Gibson, A\BPBI H.%
, Hogg, A\BPBI M.%
, Bachman, S\BPBI D.%
, Keating, S\BPBI R.%
\BCBL {}\ \BBA {} Velzeboer, N.%
\end{APACrefauthors}%
\unskip\
\newblock
\APACrefYearMonthDay{2021}{}{}.
\newblock
{\BBOQ}\APACrefatitle {A new open source implementation of {L}agrangian filtering: A method to identify internal waves in high-resolution simulations} {A new open source implementation of {L}agrangian filtering: A method to identify internal waves in high-resolution simulations}.{\BBCQ}
\newblock
\APACjournalVolNumPages{Journal of Advances in Modeling Earth Systems}{13}{10}{e2021MS002616}.
\newblock
\begin{APACrefDOI} \doi{10.1029/2021MS002616} \end{APACrefDOI}
\PrintBackRefs{\CurrentBib}

\bibitem [\protect \citeauthoryear {%
Shcherbina%
\ \protect \BOthers {.}}{%
Shcherbina%
\ \protect \BOthers {.}}{%
{\protect \APACyear {2013}}%
}]{%
shcherbina2013statistics}
\APACinsertmetastar {%
shcherbina2013statistics}%
\begin{APACrefauthors}%
Shcherbina, A\BPBI Y.%
, D'Asaro, E\BPBI A.%
, Lee, C\BPBI M.%
, Klymak, J\BPBI M.%
, Molemaker, M\BPBI J.%
\BCBL {}\ \BBA {} McWilliams, J\BPBI C.%
\end{APACrefauthors}%
\unskip\
\newblock
\APACrefYearMonthDay{2013}{}{}.
\newblock
{\BBOQ}\APACrefatitle {Statistics of vertical vorticity, divergence, and strain in a developed submesoscale turbulence field} {Statistics of vertical vorticity, divergence, and strain in a developed submesoscale turbulence field}.{\BBCQ}
\newblock
\APACjournalVolNumPages{Geophysical Research Letters}{40}{17}{4706--4711}.
\newblock
\begin{APACrefDOI} \doi{10.1002/grl.50919} \end{APACrefDOI}
\PrintBackRefs{\CurrentBib}

\bibitem [\protect \citeauthoryear {%
Sinha%
\ \BBA {} Abernathey%
}{%
Sinha%
\ \BBA {} Abernathey%
}{%
{\protect \APACyear {2021}}%
}]{%
sinha2021estimating}
\APACinsertmetastar {%
sinha2021estimating}%
\begin{APACrefauthors}%
Sinha, A.%
\BCBT {}\ \BBA {} Abernathey, R\BPBI P.%
\end{APACrefauthors}%
\unskip\
\newblock
\APACrefYearMonthDay{2021}{}{}.
\newblock
{\BBOQ}\APACrefatitle {Estimating ocean surface currents with machine learning} {Estimating ocean surface currents with machine learning}.{\BBCQ}
\newblock
\APACjournalVolNumPages{Frontiers in Marine Science}{8}{}{672477}.
\newblock
\begin{APACrefDOI} \doi{10.3389/fmars.2021.672477} \end{APACrefDOI}
\PrintBackRefs{\CurrentBib}

\bibitem [\protect \citeauthoryear {%
Smith%
, Penny%
, Platt%
\BCBL {}\ \BBA {} Chen%
}{%
Smith%
\ \protect \BOthers {.}}{%
{\protect \APACyear {2023}}%
}]{%
smith2023temporal}
\APACinsertmetastar {%
smith2023temporal}%
\begin{APACrefauthors}%
Smith, T\BPBI A.%
, Penny, S\BPBI G.%
, Platt, J\BPBI A.%
\BCBL {}\ \BBA {} Chen, T\BHBI C.%
\end{APACrefauthors}%
\unskip\
\newblock
\APACrefYearMonthDay{2023}{}{}.
\newblock
{\BBOQ}\APACrefatitle {Temporal subsampling diminishes small spatial scales in recurrent neural network emulators of geophysical turbulence} {Temporal subsampling diminishes small spatial scales in recurrent neural network emulators of geophysical turbulence}.{\BBCQ}
\newblock
\APACjournalVolNumPages{Journal of Advances in Modeling Earth Systems}{15}{12}{e2023MS003792}.
\newblock
\begin{APACrefDOI} \doi{10.1029/2023MS003792} \end{APACrefDOI}
\PrintBackRefs{\CurrentBib}

\bibitem [\protect \citeauthoryear {%
Stern%
, Uchida%
\BCBL {}\ \BBA {} Abernathey%
}{%
Stern%
\ \protect \BOthers {.}}{%
{\protect \APACyear {2022}}%
}]{%
swotadacogcm2022}
\APACinsertmetastar {%
swotadacogcm2022}%
\begin{APACrefauthors}%
Stern, C\BPBI I.%
, Uchida, T.%
\BCBL {}\ \BBA {} Abernathey, R\BPBI P.%
\end{APACrefauthors}%
\unskip\
\newblock
\APACrefYearMonthDay{2022}{}{}.
\newblock
\APACrefbtitle {{\tt swot\_adac\_ogcms}: {Documentation and notebooks for the SWOT Adopt-a-Crossover Model Intercomparison}.} {{\tt swot\_adac\_ogcms}: {Documentation and notebooks for the SWOT Adopt-a-Crossover Model Intercomparison}.}
\newblock
\begin{APACrefURL} \url{https://github.com/pangeo-data/swot_adac_ogcms} \end{APACrefURL}
\newblock
\begin{APACrefDOI} \doi{10.5281/zenodo.6762536} \end{APACrefDOI}
\PrintBackRefs{\CurrentBib}

\bibitem [\protect \citeauthoryear {%
Tchilibou%
\ \protect \BOthers {.}}{%
Tchilibou%
\ \protect \BOthers {.}}{%
{\protect \APACyear {2024}}%
}]{%
tchilibou2024internal}
\APACinsertmetastar {%
tchilibou2024internal}%
\begin{APACrefauthors}%
Tchilibou, M.%
, Carrere, L.%
, Lyard, F.%
, Ubelmann, C.%
, Dibarboure, G.%
, Zaron, E\BPBI D.%
\BCBL {}\ \BBA {} Arbic, B\BPBI K.%
\end{APACrefauthors}%
\unskip\
\newblock
\APACrefYearMonthDay{2024}{}{}.
\newblock
{\BBOQ}\APACrefatitle {{Internal tides off the Amazon shelf in the western tropical Atlantic: Analysis of SWOT Cal/Val Mission Data}} {{Internal tides off the Amazon shelf in the western tropical Atlantic: Analysis of SWOT Cal/Val Mission Data}}.{\BBCQ}
\newblock
\APACjournalVolNumPages{EGUsphere}{2024}{}{1--23}.
\newblock
\begin{APACrefDOI} \doi{10.5194/egusphere-2024-1857} \end{APACrefDOI}
\PrintBackRefs{\CurrentBib}

\bibitem [\protect \citeauthoryear {%
Torrence%
\ \BBA {} Compo%
}{%
Torrence%
\ \BBA {} Compo%
}{%
{\protect \APACyear {1998}}%
}]{%
torrence1998practical}
\APACinsertmetastar {%
torrence1998practical}%
\begin{APACrefauthors}%
Torrence, C.%
\BCBT {}\ \BBA {} Compo, G\BPBI P.%
\end{APACrefauthors}%
\unskip\
\newblock
\APACrefYearMonthDay{1998}{}{}.
\newblock
{\BBOQ}\APACrefatitle {A practical guide to wavelet analysis} {A practical guide to wavelet analysis}.{\BBCQ}
\newblock
\APACjournalVolNumPages{Bulletin of the American Meteorological society}{79}{1}{61--78}.
\newblock
\begin{APACrefDOI} \doi{10.1175/1520-0477(1998)079<0061:APGTWA>2.0.CO;2} \end{APACrefDOI}
\PrintBackRefs{\CurrentBib}

\bibitem [\protect \citeauthoryear {%
Torres%
\ \protect \BOthers {.}}{%
Torres%
\ \protect \BOthers {.}}{%
{\protect \APACyear {2018}}%
}]{%
Torres2018}
\APACinsertmetastar {%
Torres2018}%
\begin{APACrefauthors}%
Torres, H\BPBI S.%
, Klein, P.%
, Menemenlis, D.%
, Qiu, B.%
, Su, Z.%
, Wang, J.%
\BDBL {}Fu, L\BHBI L.%
\end{APACrefauthors}%
\unskip\
\newblock
\APACrefYearMonthDay{2018}{}{}.
\newblock
{\BBOQ}\APACrefatitle {{Partitioning Ocean Motions Into Balanced Motions and Internal Gravity Waves: A Modeling Study in Anticipation of Future Space Missions}} {{Partitioning Ocean Motions Into Balanced Motions and Internal Gravity Waves: A Modeling Study in Anticipation of Future Space Missions}}.{\BBCQ}
\newblock
\APACjournalVolNumPages{Journal of Geophysical Research: Oceans}{123}{11}{8084--8105}.
\newblock
\begin{APACrefDOI} \doi{10.1029/2018JC014438} \end{APACrefDOI}
\PrintBackRefs{\CurrentBib}

\bibitem [\protect \citeauthoryear {%
Uchida%
, Jamet%
, Poje%
\BCBL {}\ \BBA {} Dewar%
}{%
Uchida%
\ \protect \BOthers {.}}{%
{\protect \APACyear {2021}}%
}]{%
uchida2021ensemble}
\APACinsertmetastar {%
uchida2021ensemble}%
\begin{APACrefauthors}%
Uchida, T.%
, Jamet, Q.%
, Poje, A.%
\BCBL {}\ \BBA {} Dewar, W\BPBI K.%
\end{APACrefauthors}%
\unskip\
\newblock
\APACrefYearMonthDay{2021}{}{}.
\newblock
{\BBOQ}\APACrefatitle {An ensemble-based eddy and spectral analysis, with application to the {Gulf Stream}} {An ensemble-based eddy and spectral analysis, with application to the {Gulf Stream}}.{\BBCQ}
\newblock
\APACjournalVolNumPages{Journal of Advances in Modeling Earth Systems}{}{}{e2021MS002692}.
\newblock
\begin{APACrefDOI} \doi{10.1029/2021MS002692} \end{APACrefDOI}
\PrintBackRefs{\CurrentBib}

\bibitem [\protect \citeauthoryear {%
Uchida%
\ \protect \BOthers {.}}{%
Uchida%
\ \protect \BOthers {.}}{%
{\protect \APACyear {2022}}%
}]{%
uchida2022cloud}
\APACinsertmetastar {%
uchida2022cloud}%
\begin{APACrefauthors}%
Uchida, T.%
, Le~Sommer, J.%
, Stern, C\BPBI I.%
, Abernathey, R\BPBI P.%
, Holdgraf, C.%
, Albert, A.%
\BDBL {}others%
\end{APACrefauthors}%
\unskip\
\newblock
\APACrefYearMonthDay{2022}{}{}.
\newblock
{\BBOQ}\APACrefatitle {Cloud-based framework for inter-comparing submesoscale permitting realistic ocean models} {Cloud-based framework for inter-comparing submesoscale permitting realistic ocean models}.{\BBCQ}
\newblock
\APACjournalVolNumPages{Geoscientific Model Development}{15}{}{5829–5856}.
\newblock
\begin{APACrefDOI} \doi{10.5194/gmd-15-5829-2022} \end{APACrefDOI}
\PrintBackRefs{\CurrentBib}

\bibitem [\protect \citeauthoryear {%
Uchida%
\ \protect \BOthers {.}}{%
Uchida%
\ \protect \BOthers {.}}{%
{\protect \APACyear {2023}}%
}]{%
xrft2021}
\APACinsertmetastar {%
xrft2021}%
\begin{APACrefauthors}%
Uchida, T.%
, Rokem, A.%
, Squire, D.%
, Nicholas, T.%
, Abernathey, R\BPBI P.%
, Soler, S.%
\BDBL {}others%
\end{APACrefauthors}%
\unskip\
\newblock
\APACrefYearMonthDay{2023}{}{}.
\newblock
\APACrefbtitle {{{\tt xrft}: Fourier transforms for xarray data [Software]}.} {{{\tt xrft}: Fourier transforms for xarray data [Software]}.}
\newblock
\APAChowpublished {Zenodo}.
\newblock
\begin{APACrefURL} \url{https://xrft.readthedocs.io/en/latest/} \end{APACrefURL}
\newblock
\begin{APACrefDOI} \doi{10.5281/zenodo.1402635} \end{APACrefDOI}
\PrintBackRefs{\CurrentBib}

\bibitem [\protect \citeauthoryear {%
Vallis%
}{%
Vallis%
}{%
{\protect \APACyear {2006}}%
}]{%
Vallis:2006aa}
\APACinsertmetastar {%
Vallis:2006aa}%
\begin{APACrefauthors}%
Vallis, G.%
\end{APACrefauthors}%
\unskip\
\newblock
\APACrefYear{2006}.
\newblock
\APACrefbtitle {{Atmospheric and Oceanic Fluid Dynamics}} {{Atmospheric and Oceanic Fluid Dynamics}}.
\newblock
\APACaddressPublisher{}{Cambridge}.
\PrintBackRefs{\CurrentBib}

\bibitem [\protect \citeauthoryear {%
C.~Wang%
, Liu%
\BCBL {}\ \BBA {} Lin%
}{%
C.~Wang%
\ \protect \BOthers {.}}{%
{\protect \APACyear {2023}}%
{\protect \APACexlab {{\protect \BCnt {1}}}}}]{%
Wang_2023}
\APACinsertmetastar {%
Wang_2023}%
\begin{APACrefauthors}%
Wang, C.%
, Liu, Z.%
\BCBL {}\ \BBA {} Lin, H.%
\end{APACrefauthors}%
\unskip\
\newblock
\APACrefYearMonthDay{2023{\protect \BCnt {1}}}{}{}.
\newblock
{\BBOQ}\APACrefatitle {On Dynamical Decomposition of Multiscale Oceanic Motions} {On dynamical decomposition of multiscale oceanic motions}.{\BBCQ}
\newblock
\APACjournalVolNumPages{Journal of Advances in Modeling Earth Systems}{15}{3}{}.
\newblock
\begin{APACrefDOI} \doi{10.1029/2022MS003556} \end{APACrefDOI}
\PrintBackRefs{\CurrentBib}

\bibitem [\protect \citeauthoryear {%
C.~Wang%
, Liu%
\BCBL {}\ \BBA {} Lin%
}{%
C.~Wang%
\ \protect \BOthers {.}}{%
{\protect \APACyear {2023}}%
{\protect \APACexlab {{\protect \BCnt {2}}}}}]{%
wang2023simple}
\APACinsertmetastar {%
wang2023simple}%
\begin{APACrefauthors}%
Wang, C.%
, Liu, Z.%
\BCBL {}\ \BBA {} Lin, H.%
\end{APACrefauthors}%
\unskip\
\newblock
\APACrefYearMonthDay{2023{\protect \BCnt {2}}}{}{}.
\newblock
{\BBOQ}\APACrefatitle {A simple approach for disentangling vortical and wavy motions of oceanic flows} {A simple approach for disentangling vortical and wavy motions of oceanic flows}.{\BBCQ}
\newblock
\APACjournalVolNumPages{Journal of Physical Oceanography}{53}{5}{1237--1249}.
\newblock
\begin{APACrefDOI} \doi{10.1175/JPO-D-22-0148.1} \end{APACrefDOI}
\PrintBackRefs{\CurrentBib}

\bibitem [\protect \citeauthoryear {%
G.~Wang%
\ \protect \BOthers {.}}{%
G.~Wang%
\ \protect \BOthers {.}}{%
{\protect \APACyear {2024}}%
}]{%
wang2024indian}
\APACinsertmetastar {%
wang2024indian}%
\begin{APACrefauthors}%
Wang, G.%
, Cai, W.%
, Santoso, A.%
, Abram, N.%
, Ng, B.%
, Yang, K.%
\BDBL {}others%
\end{APACrefauthors}%
\unskip\
\newblock
\APACrefYearMonthDay{2024}{}{}.
\newblock
{\BBOQ}\APACrefatitle {The {Indian Ocean Dipole} in a warming world} {The {Indian Ocean Dipole} in a warming world}.{\BBCQ}
\newblock
\APACjournalVolNumPages{Nature Reviews Earth \& Environment}{}{}{1--17}.
\newblock
\begin{APACrefDOI} \doi{10.1038/s43017-024-00573-7} \end{APACrefDOI}
\PrintBackRefs{\CurrentBib}

\bibitem [\protect \citeauthoryear {%
Xiao%
\ \protect \BOthers {.}}{%
Xiao%
\ \protect \BOthers {.}}{%
{\protect \APACyear {2023}}%
}]{%
xiao2023reconstruction}
\APACinsertmetastar {%
xiao2023reconstruction}%
\begin{APACrefauthors}%
Xiao, Q.%
, Balwada, D.%
, Jones, C\BPBI S.%
, Herrero-Gonz{\'a}lez, M.%
, Smith, K\BPBI S.%
\BCBL {}\ \BBA {} Abernathey, R\BPBI P.%
\end{APACrefauthors}%
\unskip\
\newblock
\APACrefYearMonthDay{2023}{}{}.
\newblock
{\BBOQ}\APACrefatitle {Reconstruction of surface kinematics from sea surface height using neural networks} {Reconstruction of surface kinematics from sea surface height using neural networks}.{\BBCQ}
\newblock
\APACjournalVolNumPages{Journal of Advances in Modeling Earth Systems}{15}{10}{e2023MS003709}.
\newblock
\begin{APACrefDOI} \doi{10.1029/2023MS003709} \end{APACrefDOI}
\PrintBackRefs{\CurrentBib}

\bibitem [\protect \citeauthoryear {%
Xu%
\ \protect \BOthers {.}}{%
Xu%
\ \protect \BOthers {.}}{%
{\protect \APACyear {2022}}%
}]{%
xu2022spatial}
\APACinsertmetastar {%
xu2022spatial}%
\begin{APACrefauthors}%
Xu, X.%
, Chassignet, E\BPBI P.%
, Wallcraft, A\BPBI J.%
, Arbic, B\BPBI K.%
, Buijsman, M\BPBI C.%
\BCBL {}\ \BBA {} Solano, M.%
\end{APACrefauthors}%
\unskip\
\newblock
\APACrefYearMonthDay{2022}{}{}.
\newblock
{\BBOQ}\APACrefatitle {On the spatial variability of the mesoscale sea surface height wavenumber spectra in the {Atlantic Ocean}} {On the spatial variability of the mesoscale sea surface height wavenumber spectra in the {Atlantic Ocean}}.{\BBCQ}
\newblock
\APACjournalVolNumPages{Journal of Geophysical Research: Oceans}{127}{10}{e2022JC018769}.
\newblock
\begin{APACrefDOI} \doi{10.1029/2022JC018769} \end{APACrefDOI}
\PrintBackRefs{\CurrentBib}

\bibitem [\protect \citeauthoryear {%
Yadidya%
\ \protect \BOthers {.}}{%
Yadidya%
\ \protect \BOthers {.}}{%
{\protect \APACyear {2024}}%
}]{%
yadidya2024phase}
\APACinsertmetastar {%
yadidya2024phase}%
\begin{APACrefauthors}%
Yadidya, B.%
, Arbic, B\BPBI K.%
, Shriver, J\BPBI F.%
, Nelson, A\BPBI D.%
, Zaron, E\BPBI D.%
, Buijsman, M\BPBI C.%
\BCBL {}\ \BBA {} Thakur, R.%
\end{APACrefauthors}%
\unskip\
\newblock
\APACrefYearMonthDay{2024}{}{}.
\newblock
{\BBOQ}\APACrefatitle {Phase-accurate internal tides in a global ocean forecast model: Potential applications for nadir and wide-swath altimetry} {Phase-accurate internal tides in a global ocean forecast model: Potential applications for nadir and wide-swath altimetry}.{\BBCQ}
\newblock
\APACjournalVolNumPages{Geophysical Research Letters}{51}{4}{e2023GL107232}.
\newblock
\begin{APACrefDOI} \doi{10.1029/2023GL107232} \end{APACrefDOI}
\PrintBackRefs{\CurrentBib}

\end{thebibliography}
%




%
%
%
%
%

\end{document}